\documentclass[fleqn,usenatbib]{mnras}

\usepackage{newtxtext,newtxmath}
\usepackage{graphicx}
\usepackage{amsmath} 
\usepackage{gensymb}
\usepackage{tabularx}
\usepackage{enumitem}
\usepackage{float}

\usepackage[T1]{fontenc}
\DeclareRobustCommand{\VAN}[3]{#2}
\let\VANthebibliography\thebibliography
\def\thebibliography{\DeclareRobustCommand{\VAN}[3]{##3}\VANthebibliography}

\title[Microlensed Accretion Discs]{Resolving the Vicinity of Supermassive Black Holes with Gravitational Microlensing}

\author[H. Best et al.]{
Henry Best,$^{1, 2, 3}$\thanks{E-mail: hbest@gradcenter.cuny.edu}
Joshua Fagin,$^{1, 2, 3}$
Georgios Vernardos,$^{2, 3, 4}$
and Matthew O'Dowd,$^{1, 2, 3}$ 
\\
$^{1}$Department of Physics, The Graduate Center of the City University of New York, 365 Fifth Avenue, New York, NY 10016, USA\\
$^{2}$Department of Astrophysics, American Museum of Natural History, Central Park West and 79th Street, NY 10024-5192, USA\\
$^{3}$Department of Physics and Astronomy, Lehman College of the CUNY, Bronx, NY 10468, USA\\
$^{4}$Institute of Physics, Laboratory of Astrophysics, Ecole Polytechnique Fédérale de Lausanne (EPFL), Observatoire de Sauverny, 1290 Versoix, Switzerland\\
\\
}

\date{Accepted XXX. Received YYY; in original form ZZZ}

\pubyear{2022}

\begin{document}
\label{firstpage}
\pagerange{\pageref{firstpage}--\pageref{lastpage}}
\maketitle

\begin{abstract}
Upcoming wide field surveys will discover thousands of new strongly lensed quasars which will be monitored with unprecedented cadence by the Legacy Survey of Space and Time (LSST). 
Many of these quasars will undergo caustic-crossing events over the 10-year LSST survey, during which the quasar's inner accretion disc crosses a caustic feature produced by an ensemble of microlenses.
Such caustic-crossing events offer the unique opportunity to probe the vicinity of the central supermassive black hole, especially when combined with high cadence, multi-instrument follow-up triggered by LSST monitoring.
To simulate the high cadence optical monitoring of caustic-crossing events, we use relativistic accretion disc models which leads to strong asymmetric features.
We develop analysis methods to measure the Innermost Stable Circular Orbit (ISCO) crossing time of isolated caustic-crossing events and benchmark their performance on our simulations. 
We also use our simulations to train a Convolutional Neural Network (CNN) to infer the black hole mass, inclination angle, and impact angle directly from these light curves.
As a pilot application of our methods, we used archival caustic-crossings of QSO 2237+0305 to estimate the black hole mass and inclination angle.
From these data, two of these methods called the second derivative and wavelet methods measure an ISCO crossing time of 48.5 and 49.5 days, corresponding to a Kerr black hole mass of $M_{\rm{BH}} = (1.5 \pm 1.2) \times 10^{9} M_{\odot}$ and $M_{\rm{BH}}  = (1.5 \pm 1.3) \times 10^{9} M_{\odot}$ respectively.
The CNN inferred $\log_{10} (M_{\rm{BH}} / M_{\odot}) = 8.35 \pm 0.30$ when trained on Schwarzschild black hole simulations, and a moderate inclination of $i = 45 \pm 23\degree$. 
These measurements are found to be consistent with previous estimates.
\end{abstract}


\section{Introduction}
\label{Introductionsection}

Active Galactic Nuclei (AGN) play an integral role in the evolution of galaxies and are important probes of the distant Universe.
It is widely accepted that they are powered by the conversion of gravitational potential energy through accretion on to a Super Massive Black Hole ~\citep[SMBH,][]{Salpeter64,Zeldovich64}.
This energy is then reprocessed through a variety of mechanisms into radiation that spans the electromagnetic spectrum~\citep[see][for a review]{Padovani17}.
Especially luminous AGNs with unobscured accretion discs are known as quasars, and are visible even at extreme cosmological distances~\citep[$z>7$, e.g.][]{Mortlock11, Banados18}. 
The accretion disc has an angular scale of nano- to micro-arcsec ($\approx$1 light-day in physical length), and therefore cannot be spatially resolved by direct imaging.
The region in the vicinity\footnote{Here we refer to the innermost few gravitational radii as the SMBH vicinity.} and under the direct influence of the SMBH is two orders of magnitude smaller than the optical disc and is likely to remain inaccessible to direct observational methods.
 
There are currently five methods that have the potential to probing the vicinity of SMBHs--two of which have been successfully realized.
The Event Horizon Telescope has recently yielded our first images of the vicinity of two accreting SMBH, M87 and Sgr A$*$ \citep[][]{Horizon19,Horizon22} using interferometry at 1.3 mm.
Very Long Baseline Interferometry on GRAVITY~\citep{Gravity17} is sensitive to relativistic effects of our Galaxy's SMBH on closely orbiting stars ~\citep[Especially S2 orbiting around Sgr A$*$, ][]{GRAVITY19, GRAVITY20} and has the potential to spatially resolve sizes down to the near infrared accretion disc in some cases, though this has not been used to constrain size scales of the vicinty of the SMBH.
Reverberation mapping of the Doppler boosted iron K$\alpha$ emission line~\citep{Zoghbi11a, Zoghbi11b, prince22, Lucchini23} and continuum lags~\citep{Kammoun21, Jha22, Guo22} have been successfully used to probe kinematics and geometry of the accretion disc down to several gravitational radii~\citep[see][for a review]{Cackett21}.
Local transit events, both by the local stellar population~\citep{Beky13} or by self-lensing flares in a binary system~\citep{Davelaar22} may also probe the central region around the black hole, but this has yet to place constraints on the vicinity of the SMBH.

The fifth approach which we investigate in this work offers the potential to scan the accretion disc of \emph{strongly lensed} quasars down to the scale of the gravitational radius.
Strong gravitational lensing of a quasar by a galaxy results in multiple, resolvable images of the quasar.
Each image may then be influenced by microlensing, gravitational lensing by the substructure of the lensing galaxy~\citep[~\citet{Chang79}, see][for a review]{Vernardos23}.
This can result in a magnification or demagnification of observed flux in the image experiencing microlensing.
The strength of a microlensing event depends on the size of the object in the \emph{source plane} (e.g. the plane of the quasar or other object of interest), such that smaller objects experience a greater impact. 
Microlensing is time-dependent due to the relative transverse motion of microlenses, source, and observer.
We observe this as wavelength-dependent flux variations over typical timescales of months to years. 

Microlensing has the ability to harness the natural magnification boost in strongly lensed systems and has already been used to study quasars.
Outside the high-magnification regime of a caustic-crossing event, microlensing is relatively insensitive to the small scale features and detailed structure of the brightness profile.
It instead depends on the overall size of the disc \citep[the half-light radius,][]{Mortonson05} relative to the Einstein Radius. 
As such, many studies use simplified profile shapes such as a two-dimensional Gaussian, with an appropriate scaling of half-light radius to observed wavelength to approximate a thin disc thermal profile~\citep{Grieger88, Agol99, Wyithe02, Bate08, Jimenez14, tomozeiu18, Bate18}.
This analysis of microlensing light curves has led to constraints on sizes and geometries of quasar emission regions, from the accretion disc \citep[e.g.][]{Pooley07, Morgan10, Blackburne11, Jimenez12, Munoz16} to the Broad Emission Line Region (BELR) \citep[e.g.][]{Schneider90, Hutsemekers94, Lewis98, Abajas02, Sluse07, ODowd11, paic22, Williams21}.
However, microlensing also has the potential to resolve the vicinity of the SMBH down to the gravitational radius, $R_{\rm g} \equiv GM_{\rm{BH}}/c^{2}$.
This power comes from caustic folds and cusps--regions in the source plane where the magnification formally diverges.
In practice, these areas produce magnifications up to $\mathcal{O} (10^{3})$ over nano-arcsec regions in the source plane for an appropriately small source (e.g. an accretion disc seen in ultra violet or X-ray wavelengths).
These \emph{caustic-crossing events} are rare, $\approx 1$ per decade per system \citep{Mosquera11}, and unfold quickly within a few weeks or months \citep[e.g.][]{Neira20}.

In the next few years, caustic-crossing events are expected to change from a rare observation into routinely observed phenomena.
This is largely due to wide field surveys, such as the Legacy Survey in Space and Time (LSST) to be carried out at the Vera C. Rubin Observatory~\citep{LSST19}.
In concert with Euclid, LSST will discover several thousand strongly-lensed quasars~\citep{Oguri2010} and monitor them in optical filters with a cadence of a few days.
Over its 10-year survey, approximately 5 per cent of these lensed quasars are expected to undergo caustic-crossing events~\citep{Mosquera11}, resulting in upwards of 300 events per year~\citep{Neira20}.
Routine LSST monitoring will detect the $\sim$ weeks-long onset of these caustic-crossing events, enabling high-cadence, multi-wavelength follow-up of the event itself.

There are two approaches that will make use of these caustic-crossing events to probe the vicinity of the black hole.
The first method resides in the spectral domain--the "g-distribution" method is an approach introduced by~\citet{Chartas17} and is used to determine various black hole parameters~\citep[see also][]{Ledvina18}.
This method analyses the relativistic Fe K$\alpha$ line's observed spectral energy distribution and places constraints on the inclination and ISCO size based on the impact of microlensing.
Spectral X-ray data is required for this analysis, and the inner accretion disc must be on or near a caustic to provide the strongest constraints.
The second method focuses on the time domain, where the analysis of caustic-crossing event light curves can probe the innermost regions of the accretion disc as introduced by~\citet{Abolmasov12a} ~\citep[see also][]{Mediavilla15b}.
The detailed structure of the disc near the black hole will imprint itself on the observed light curve during a caustic-crossing event. 
This imprint is referred to as \emph{fine structure} and is expected to contain information about the asymmetry between approaching and receding sides of the accretion disc, as well as the signature of the dark region within the ISCO.

Of particular interest regarding caustic-crossing events is the system QSO 2237+0305, also known as the Einstein Cross. 
This is a quadruply lensed quasar characterized by a lensing galaxy (Huchra's lens) at redshift $z_{\rm{l}}$ = 0.039 and a source redshift of $z_{\rm{s}}$ = 1.695~\citep{Huchra85}.
The quasar has a very high effective velocity due to its exceptionally low value of $z_{\rm{l}}$, which leads to a greater rate of caustic-crossings compared to other systems.
This system experienced the first observed microlensing event as presented in~\citet{Irwin89}, where one image of QSO 2237+0305 experienced a significant change in brightness.
Later on, \citet{Mediavilla15b} identified the possible signature of the ``black hole shadow'': The low surface brightness region expected to exist within the ISCO.
However no event has yet been observed with sufficient cadence, depth, and wavelength coverage to place strong constraints on this or other properties of the strongly curved spacetime around the SMBH.

Various authors have tried to explain the shape of high magnification events in QSO 2237+0305.
\citet{Abolmasov12a} used a relativistic thin disc and included a Doppler boost modelled by frequency shifting.
\citet{Mediavilla15b} used both a classical and relativistic thin disc in order to model three caustic-crossing event candidates observed in QSO 2237+0305.
Their relativistic model includes the effect of beaming alongside a Doppler shift and a gravitational redshift, which results in an improved fit to the observations over the non-relativistic model.
\citet{tomozeiu18} adopted a simpler approach using a crescent shaped accretion disc model to approximate a simulation of a caustic-crossing event in M87.
Each of these studies includes relativistic effects in a different way, but they do not include a full relativistic treatment of the central SMBH that may have a significant impact on the innermost regions of the accretion disc.

Quasar microlensing events have traditionally been modelled using Markov Chain Monte Carlo (MCMC) to best fit the light curves~\citep[e.g.][]{Yonehara01, Kochanek04}.
More recently, machine learning has become a common tool applied to strongly lensed quasars, from source finding and classification in surveys~\citep{Busca18, Guo19} to estimating redshifts and velocities~\citep{Pasquet18, Rastegar22}.
Deep learning has also been used to model quasar variability~\citep{Quasar_variability_ML, Fagin23}.
In the case of microlensing, neural networks have been found to estimate half-light radii of accretion discs without needing particular information of the temperature gradient~\citep{Vernardos_2019}.

In this work, we build up simulations of caustic-crossing events that employ a realistic analytic model of the inner accretion disc and account for special and general relativistic effects (e.g. Doppler shifting of wavelengths, Doppler beaming, and local lensing through geodesic tracing around the central SMBH).
From these, we simulate intensive optical follow-up to assess our ability to infer a range of parameters of the inner disc using the fine structure of caustic-crossing events.
In Section~\ref{Modelsection} we describe the accretion disc and microlensing models that we use and present the resulting features of our simulated light curves. 
In Section~\ref{Analysissection}, we develop and apply three different analysis methods to determine if we may extract physical parameters from the simulated caustic-crossing events.
The results for our simulations are presented in Section~\ref{resultssection}, where we compare the recovered ISCO size of the light curve fitting methods with the predicted values from our machine learning method.
We apply our developed methods to previous observations of a caustic-crossing event in QSO 2237+0305 in Section~\ref{Results2237}.
In Section~\ref{Conclusionsection}, we present our conclusions on these methods and their ability to place constraints on the mass and inclination of QSO 2237+0305's SMBH.
Within this work, we assume a flat $\Lambda$CDM cosmology with Hubble constant 70 km s$^{-1}$ Mpc$^{-1}$ and $\Omega_{0}$~=~0.3.

\section{Simulated light curves}
\label{Modelsection}
In this section, we detail how we simulate light curves of caustic-crossing events.
This is done by taking the convolution of a model surface brightness profile of the accretion disc (Section~\ref{QModelsection}) with a simulated microlensing magnification map (Section~\ref{LModelsection}).
The cadence of observations and photometric noise is modelled according to an intensive optical follow-up observation of the caustic-crossing event (Section~\ref{LCSubSection}).

\subsection{Accretion disc model}
\label{QModelsection}

\subsubsection{The thin disc model}
\label{AdiscSection}

Our starting point is the standard thin disc model~\citep{ShakuraSunyaev} to which we add the following relativistic effects: Doppler shifting of the continuum, Doppler beaming, and gravitational lensing by the local SMBH.
It is important to distinguish local strong lensing by the central black hole from the typical strong lensing that requires a galaxy-sized lens located between the observer and the source.
This local strong lensing effect will take place in all quasars.
The temperature profile of a thin disc is defined as:
\begin{equation}
    \label{eq:disc_T}
    T(r) = \left[\frac{GM_{\rm BH}\dot{M}_{\rm BH}}{8\pi \sigma r^{3}} \left(1 - \sqrt{\frac{R_{\rm in}}{r}}\right)\right]^{1/4} ,
\end{equation}
where $M_{\rm BH}$ is the SMBH mass, $G$ and $\sigma$ are the gravitational and Stefan-Boltzmann constants, and $R_{\rm in}$ is the inner boundary of the disc.
We take $R_{\rm{in}}$ to be the ISCO: $R_{\rm in}=\alpha GM_{\rm BH}/c^2$, where $\alpha$ depends on the magnitude of the SMBH's spin ($\alpha_{*}$) and the orientation of the accretion disc's orbits relative to the SMBH's spin ($\alpha=$ 6, 1, or 9 for non-rotating, maximal SMBH spin with prograde orbits, and maximal SMBH spin with retrograde orbits respectively). 
The accretion rate, $\dot{M}_{\rm BH}$, can be expressed in terms of the Eddington ratio $R_{\rm Edd}=L/L_{\rm Edd}$, where $L = \eta \dot{M}_{\rm{BH}} c^{2}$ and $L_{\rm Edd} = 4 \pi G M_{\rm BH} m_{p} c/\sigma_{\rm T}$, which is the theoretical accretion limit due to radiation pressure.
Within these equations, $\eta$ is the conversion efficiency for gravitational energy to thermal energy, $m_{p}$ is the mass of a proton, and $\sigma_{\rm T}$ is the Thompson cross section.

Using Equation (\ref{eq:disc_T}) and assuming the disc radiates as a black body, we calculate the flux of the disc as a function of radius and (rest) wavelength.
We assume that the region interior to the ISCO does not emit any photons that reach the observer.
By taking the inclination angle $i$ into account ($0\degree$ is face on), we project the accretion disc on to the source plane by following the geodesics of photons around the central SMBH to obtain the two-dimensional projected surface brightness profile of a thin disc.

\begin{figure}
    \centering
    \includegraphics[width=0.45\textwidth]{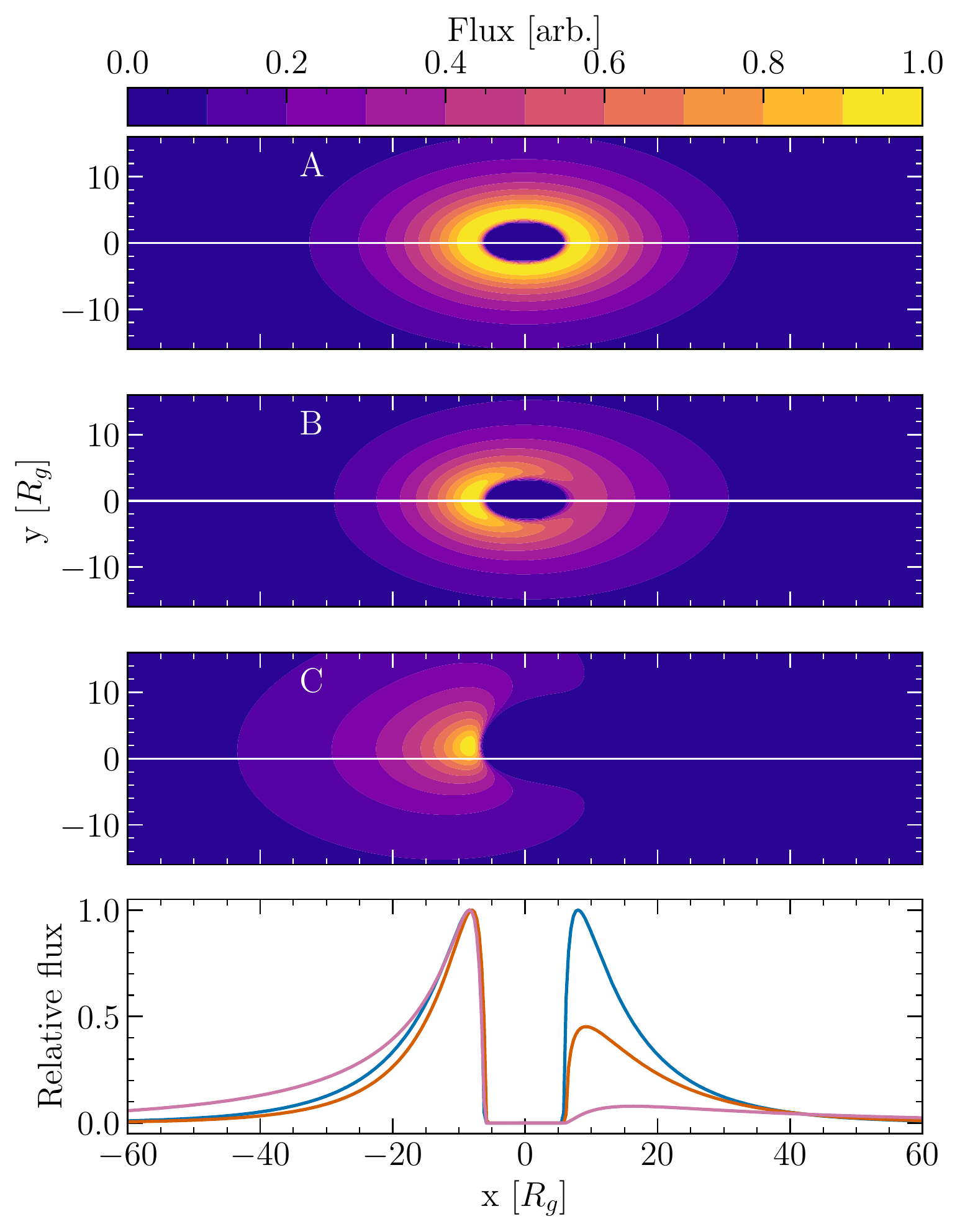}
    \caption{Normalized projected surface brightness profiles of an inclined accretion disc ($i = 60\degree$). Contours are placed at every 10 per cent interval of the maximum flux. We assume the accretion disc rotates counter-clockwise, such that the left side approaches the viewer. Panel A) The thin disc profile is symmetric before any corrections are added. Panel B) By considering Doppler shifts, a low level of asymmetry is introduced. Panel C) Ray tracing and beaming effects around the SMBH introduce a drastic increase in asymmetry between the approaching and receding sides. The bottom panel shows the relative brightness for each panel A, B, and C along the y = 0 line in blue, orange, and purple respectively.}
    \label{Figure1}
\end{figure}  

\subsubsection{Relativistic extensions}
We extend the exponential thin disc model to include the following relativistic effects: Doppler shifting, Doppler beaming, and relativistic ray tracing of geodesics. 
For each pixel on the source plane, we reverse the cosmologic and relativistic redshift contributions to find the rest wavelength which is expected to become shifted to the observed wavelength.
At these rest wavelengths, we calculate the intensity of black body radiation coming from the thin disc model. 
We apply Doppler beaming and energy shifting due to the difference between rest and observer frames.
Ray tracing and relativistic redshifting are simulated with the general relativistic ray tracing code \texttt{GYOTO}\footnote{\url{https://gyoto.obspm.fr}}~\citep{Vincent11} through an appropriate Kerr metric~\citep{Kerr63}, leading to the final projected surface brightness distribution.  
This primarily impacts the innermost regions of the disc where light bending effects are the greatest.

We illustrate the result of including each effect in Fig.~\ref{Figure1}.
Prior to the introduction of relativistic effects and ray tracing, the disc appears symmetric.
The introduction of Doppler shifting results in the brightening (dimming) for line-of-sight velocities towards (away from) the observer.
This leads to a factor of $\sim2$ asymmetry in surface brightness between the approaching and receding sides of the inner disc. 
The inclusion of geodesic tracing gives a significant enhancement of the surface brightness asymmetry from relativistic beaming---a factor of $\sim10$ for the case of $i=60\degree$.
The peak flux of the surface brightness profile and major axis of the ISCO is no longer found on the central axis, but appears to "creep" up in the source plane.

\begin{figure}
    \centering  
    \includegraphics[width=0.46\textwidth]{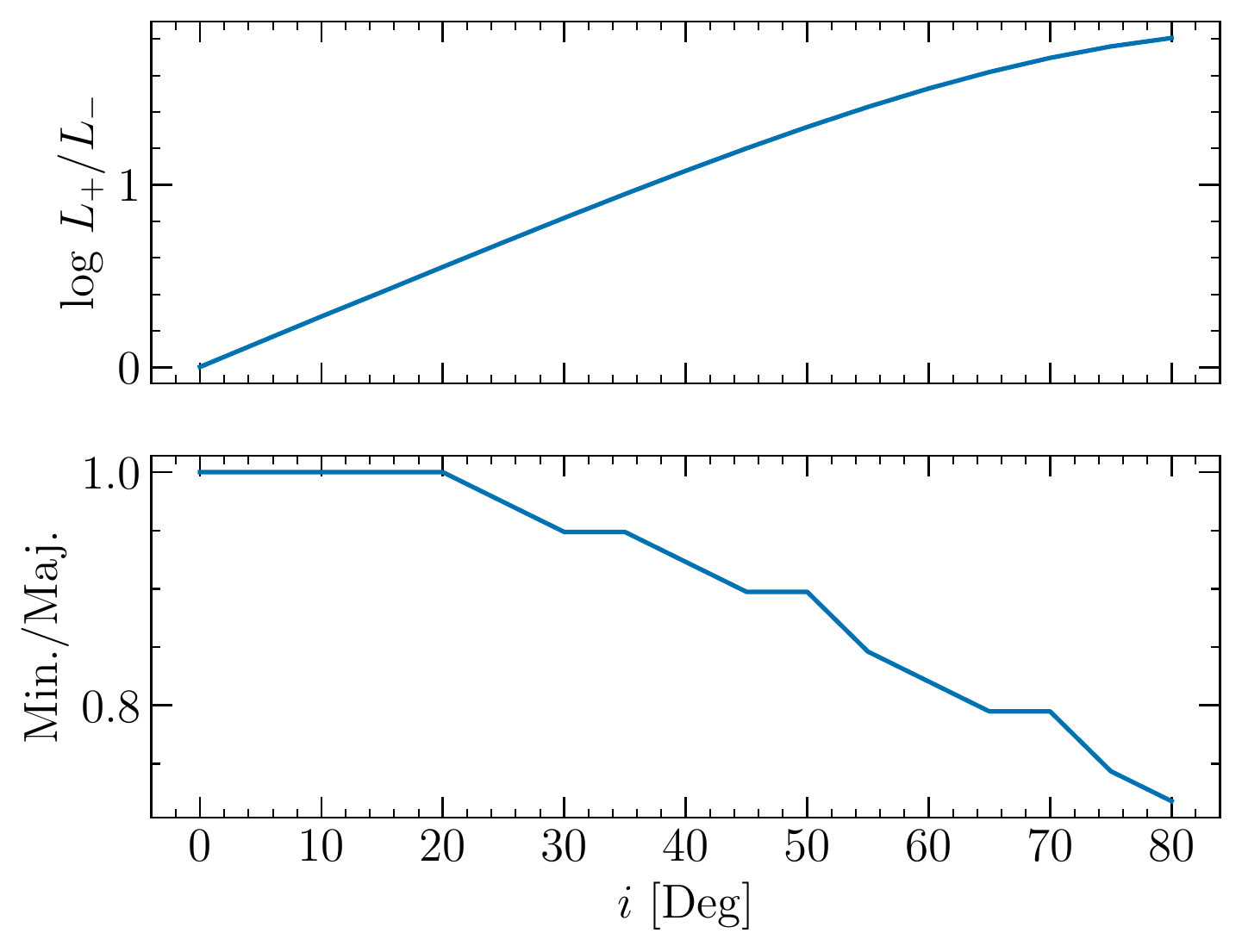}
    \caption{Effect of the inclination angle $i$ on the accretion disc asymmetry. Top: Flux asymmetry as a function of $i$ between points at $7 R_{\rm g}$ on the approaching and receding sides of the disc. These points are marked by the small crosses in Fig.~\ref{Figure3}. Bottom: The projected ISCO compression as a function of $i$. Compression is defined as the ratio of minor to major ISCO axes.}
    \label{Figure2}
\end{figure}

This asymmetry in brightness between the approaching and receding sides of the disc depends on the inclination angle.
Fig.~\ref{Figure2} demonstrates this in the top panel, as the asymmetry increases greatly as we approach an edge-on orientation.
Edge-on discs are thought to be obscured by dusty tori surrounding them, so we do not expect such extreme inclinations for quasars.

\begin{figure}
    \centering
    \includegraphics[width=0.46\textwidth]{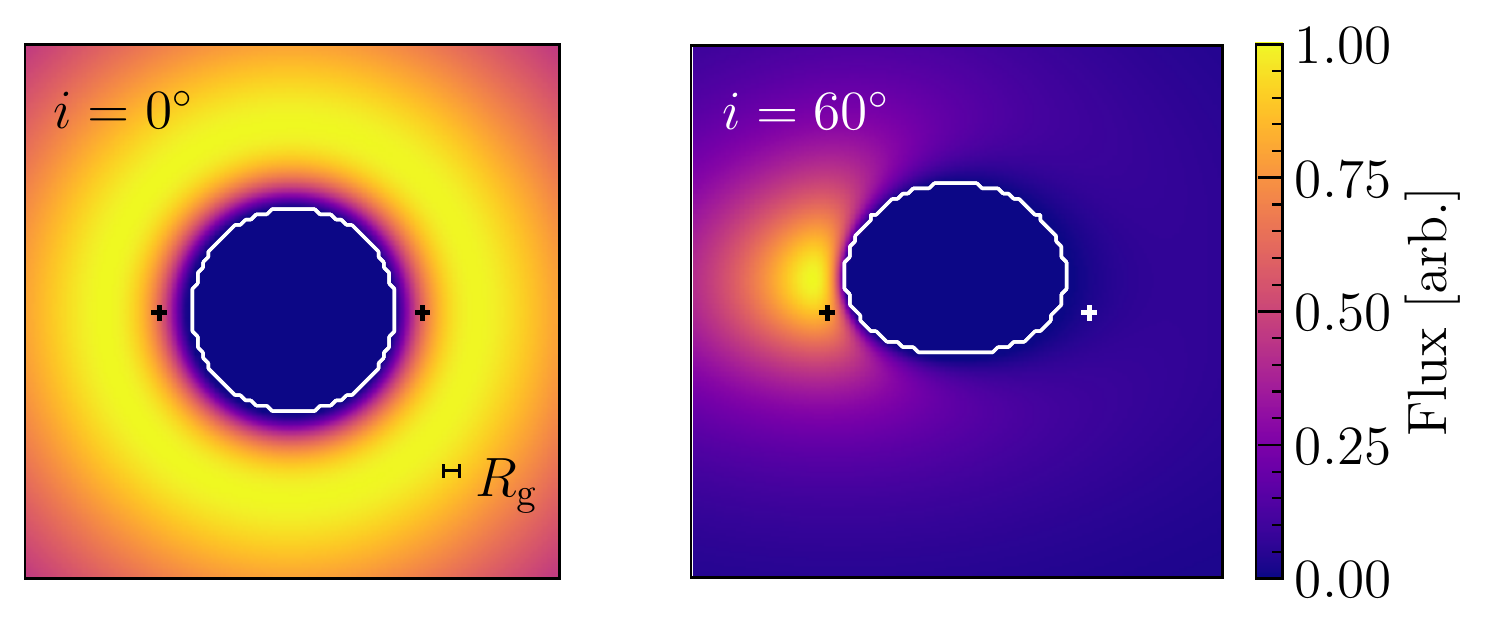}
    \caption{Central region of the accretion disc for $i = 0\degree ~\text{and}~ 60\degree$. The approaching side of the accretion disc is on the left. Crosses mark the locations used to calculate the asymmetry ratio shown in Fig.~\ref{Figure2}. The small horizontal bar overlaid on the left image indicates $1 R_{\rm g}$. A white contour is placed at the ISCO for clarity. The colour bar denotes the relative flux from the accretion disc, scaled to the peak emission. The location of the ISCO's major axis ``creeps'' up at higher inclination angles due to the strongly curved geodesics around the SMBH.}
    \label{Figure3}
\end{figure}

Tilted accretion discs may appear compressed along one axis due to the geometric projection into the source plane.
However, due to light bending effects around the SMBH, the ISCO will not experience the same apparent compression as the optical accretion disc.
Only the ISCO edge nearest to the observer appears contracted after its projection into the source plane.
This is apparent in the bottom panel of Fig.~\ref{Figure2}, where the maximum compression approaches $\sim 70$ per cent at $i = 80\degree$, and in the right panel of Fig.~\ref{Figure3} where the near side of the accretion disc is at the bottom of the image.

\subsection{Microlensing magnification}
\label{LModelsection}
The caustic network produced by an ensemble of stellar-mass, compact objects depends on the local values of the convergence and shear within the lensing galaxy at the positions of the quasar images.
These networks can vary from isolated diamond-shaped caustics to dense networks of overlapping and nested ones ~\citep[see Fig. 12 in][or the GERLUMPH\footnote{\url{https://gerlumph.swin.edu.au/status/}} database for examples]{Vernardos23}.
Despite this, the basic property of a caustic and its analytical description remain unchanged \citep{Fluke99}.
This means that if a source is smaller than the distance between two consecutive caustics, then all caustic-crossing events can be treated as isolated.
Here we use an isolated caustic to produce our mock light curves, expecting our results to be applicable also in cases of more complicated caustic structures as long as the source is small enough.
The main length scale that sets the size of a caustic is the Einstein radius $R_{\rm E}$, defined as:
\begin{equation}
    R_{\rm E} = \sqrt{\frac{4GM_{\text{l}}}{c^{2}} \frac{D_{\text{s}} D_{\text{ls}}}{D_{\text{l}}}}
    \label{R_einstein}
\end{equation}
where $M_{\text{l}}$ is the mass of the microlens and $D_{\text{l}}~ D_{\text{s}} ~D_{\text{ls}}$ are the angular diameter distances between the observer, lens, and source.
Any distance with one index is implied to be with respect to the observer.

\begin{figure}
    \centering
    \includegraphics[width=0.47\textwidth]{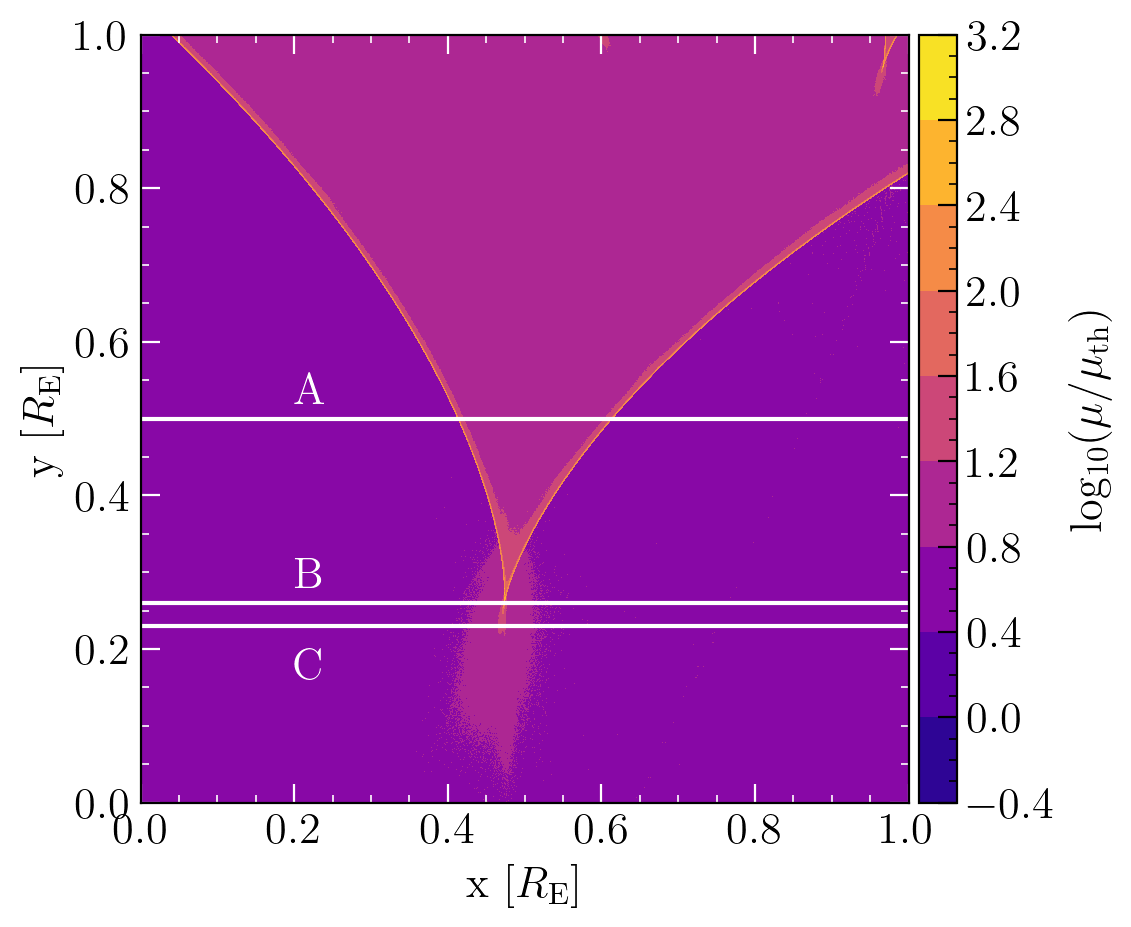}
    \caption{Magnification map of an isolated caustic used in simulating caustic-crossing events within this work. Example light curve tracks are shown for a fold (A), passing just inside the cusp (B), and for passing outside the cusp (C). The colour bar denotes the logarithm of the magnification with respect to the theoretical magnification.}
    \label{Figure4}
 \end{figure}

Fig.~\ref{Figure4} shows the high-resolution, zoomed-in magnification map around an isolated caustic of a GERLUMPH map created using the \texttt{GPU-D}\footnote{\url{https://gerlumph.swin.edu.au/software/}} code \citep{Thompson10} which was used in this study.
The map was 1.0 $R_{\rm E}$ on each side, and the resolution $10^{4}$ pixels per side.
The compact and smooth convergence components were 0.328 and 0.082 respectively, and the shear was 0.38.
Magnification of this map at each pixel was calculated as done in ~\cite{Vernardos14}.

\begin{figure}
    \centering
    \includegraphics[width=0.47\textwidth]{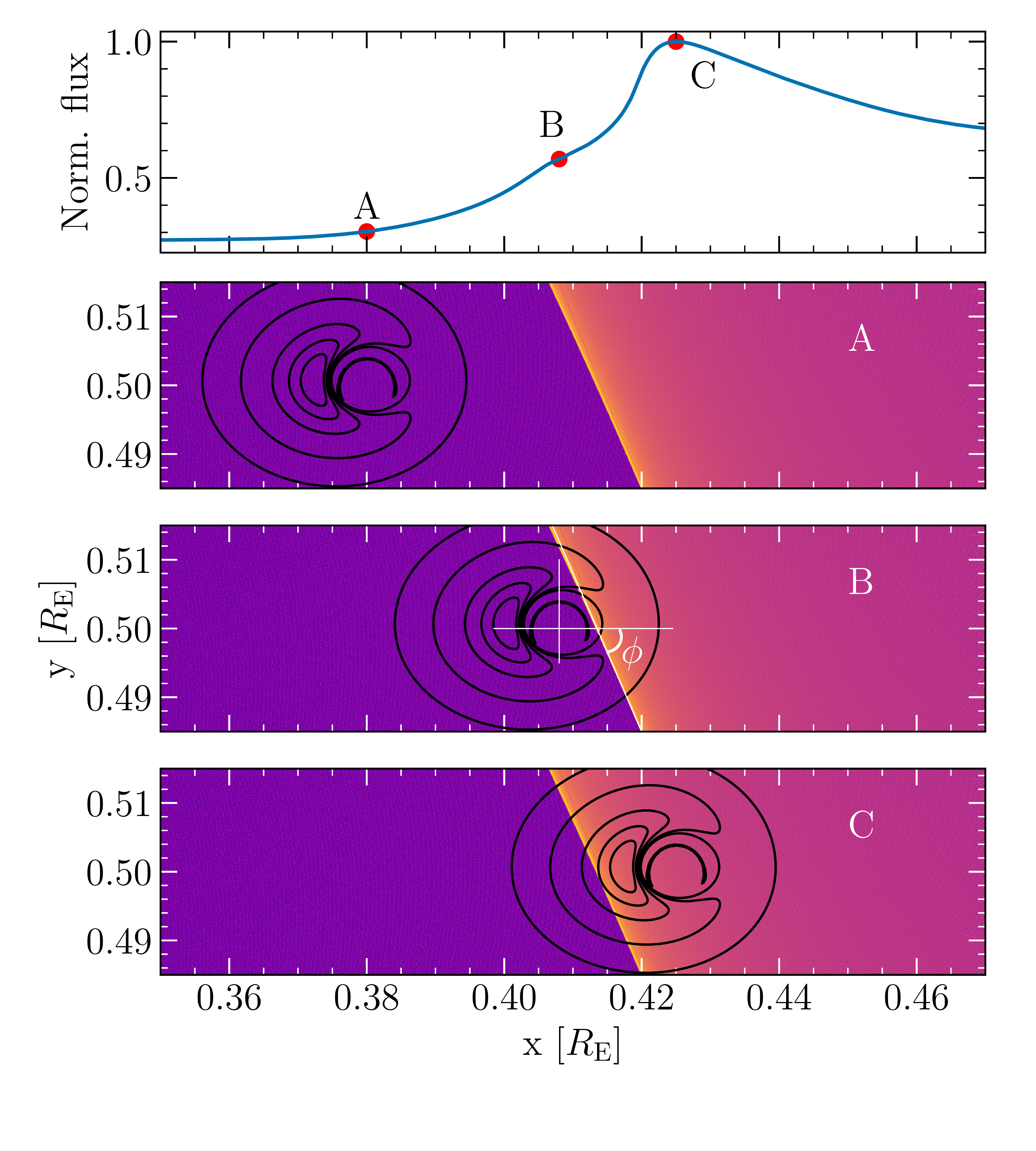}
    \caption{Top: Sample light curve of a caustic-crossing event for a moderately inclined accretion disc ($i = 50\degree$). The labeled positions in the top panel are shown in each subsequent panel, highlighting certain points in the caustic-crossing event. Contour levels of the disc are placed at 0.1, 0.2, 0.4, 0.6, and 0.8 of the peak flux at 600 nm. Position A represents the initial rise of the event. Position B represents the start of the ISCO crossing event. Position C represents the peak of the caustic-crossing event. In panel B, we highlight the major and minor axes of the accretion disc, the caustic, and the impact angle $\phi$.}
    \label{Figure5}
\end{figure} 

\subsection{Simulating light curves}
\label{LCSubSection}
The free parameters of our accretion disc model are the SMBH mass and spin, the Eddington ratio, and the inclination angle.
The lensing parameters consist of the impact angle (the disc orientation relative to the caustic fold), the redshifts of the lens and source, and the magnification map parameters.
The magnification map parameters are less important due to the universal behaviour of an isolated caustic-crossing event.
Degeneracies between some of these parameters are significant, such as how $M_{\text{BH}}$ and spin are degenerate with respect to the ISCO radius. 
When focused on the ISCO radius alone, spin may mimic up to one decade of $M_{\text{BH}}$.
There is another important degeneracy in microlensing: the magnification of a source depends on its relative size with respect to the Einstein radius of the microlenses.
This relative size is what determines the scale of fine structure on the resulting light curves during a caustic-crossing event.
Therefore, all parameters that go into calculating $R_{\rm E}$ (Equation~\ref{R_einstein}) and $R_{\rm S}$ (Equation~\ref{R_source}) are degenerate to some degree.
We detail our choices of all parameters in this section and note that the parameter space explored is not intended to be exhaustive, but rather to allow us to explore the performance of our measurements for a wide range of the plausible parameter space.

For the SMBH mass, we use the range $7.7 \leq $ log$_{10} (M_{\rm BH} / M_{\odot}) \leq  9.7$, which is a typical range for quasars with dynamically measured masses~\citep{McConnell13}.
The ISCO radius also depends on the SMBH spin and is expected to be between 1 and 9 $R_{\rm{g}}$.
In this work, we do not explore the range of spin, but rather assume the case of a non-rotating black hole with ISCO radius $6R_{\rm{g}}$.
Hence any inference we make of black hole mass from the ISCO radius implies a $\sim 1$ dex uncertainty.
We assume an accretion rate yielding an Eddington ratio of $R_{\rm{Edd}} = 0.15$, chosen to represent a typical quasar at $z\sim1$~\citep{Kelly10}.
The thin disc model is not expected to apply for significantly higher Eddington ratios where other accretion models are predicted (e.g. ADAFs, slim discs, etc).
We set $z_{\text{l}} = 0.5$ for all of our simulations and explore source redshifts of $1 \leq z_{\text{s}} \leq 3$.
This corresponds to the peak of expected lens redshifts and spans a wide range of source redshifts expected of lensed quasar systems to be discovered by LSST~\citep{Oguri2010}.
Finally we explore inclination angles with $i \leq 80\degree$, assuming that larger inclinations (more edge-on discs) are obscured by the dusty torus~\citep{Antonucci93}.

When simulating the magnification due to microlensing, the most important parameters are the Einstein radius and the source size.
The theoretical source size is calculated as in~\cite{Morgan10} and~\cite{Mosquera11}:
\begin{equation}
    \label{R_source}
    R_{\rm S}(\lambda) = 9.7 \times 10^{13} \left( \frac{\lambda}{\mu m} \right)^{4/3} \left( \frac{M_{\rm BH}}{10^{9}M_{\odot}}\right)^{2/3} \left( \frac{R_{\rm edd}}{\eta} \right)^{1/3} \text{m}
\end{equation}
where lambda is the (rest) wavelength in $\mu$m, and $\eta$ is the efficiency of the conversion of gravitational potential into thermal energy.
We assumed an efficiency value of $\eta = 0.1$.
Our accretion discs are simulated at observed wavelengths 477, 623, 762, 913 nm to represent the effective wavelengths of the $g$, $r$, $i$, and $z$ filters.
We consider the cosmological redshift at $z_{\rm{s}}$, leading to rest frame wavelengths ranging from 119 to 457 nm.
The Einstein radius of microlens(es) is calculated using Equation (~\ref{R_einstein}).

\cite{Mosquera11} describe how the relative size scale of the disc ($R_{\rm S}$) with respect to the Einstein radius ($R_{\rm E}$) is what determines the impact of microlensing. 
Within their sample of lensed quasars, this ratio was determined to be $0.001 < R_{\rm S} / R_{\rm E} < 0.25$.
When we compared our relative source sizes to this sample, we found that our relative size scales fall between $0.0002 < R_{\rm S} / R_{\rm E} < 0.11$.
In our simulation we considered the source redshift, leading to systematically smaller accretion discs.
With this in mind, the parameter range we use is a reasonable representation of the strongly lensed quasars expected to be discovered in wide-field surveys.

As shown in Fig.~\ref{Figure2}, the inclination angle plays two major roles in caustic-crossing events.
First, the accretion disc's brightness profile becomes asymmetric as inclination is increased, leading to the amplification of the approaching side and the suppression of the receding side.
This asymmetry can have a significant impact on caustic-crossing event light curves as shown in Fig.~\ref{Figure5}.
Second, the inclination may cause a reduction in the projected ISCO crossing length $L_{\rm{ISCO}}(M_{\rm{BH}}, i, \phi)$.
We define the \emph{impact angle} of $\phi = 0\degree$ to correspond to the case where the disc's major axis is parallel to the caustic fold in the source plane.
Assuming motion is perpendicular to the caustic fold, the accretion disc is scanned along its minor axis. 
This gives us the greatest reduction of $L_{\rm{ISCO}}$ and the least impact due to the disc's asymmetry (e.g. both approaching and receding sides are amplified at the same time).
An impact angle of $\phi = + 90\degree$ corresponds to the receding side of the accretion disc crossing the caustic first, while an angle of $\phi = - 90\degree$ corresponds to the approaching side of the accretion disc crossing first.
Neither $\phi = \pm 90\degree$ cause compression of $L_{\rm{ISCO}}$, as the major axis is being swept by the caustic.
Intermediate values of $\phi$ determine the precise value of $L_{\rm{ISCO}}$.

We calculate our signal-to-noise ratio using the Gemini South Integration Time Calculator for GMOS imaging, assuming 1 hour exposures in the targeted optical bands and with best observing conditions.
This corresponds to the best 20th percentile viewing conditions for ground based follow-up of an imminent caustic-crossing event.
Scatter is applied to simulate photometric uncertainty assuming Poisson noise.
We acknowledge that any given caustic-crossing follow-up may experience a range of observing conditions.
However with 1000's of potential events over the LSST survey, some will enjoy excellent follow-up conditions. 
For this study we use the best-case scenario to demonstrate the potential of our methods.

The cadence of simulated observations is set in terms of the Einstein radius because the actual transverse velocity of the disc across the magnification map can vary greatly (see \citet{Neira20} for examples).
We simulate evenly spaced observations such that each time interval between observations has a length of $10^{-4} R_{\rm E}$.
The total length of each light curve is set to be 0.15 $R_{\text{E}}$ in the source plane in order to assure the inner accretion disc fully crosses the caustic.
In physical terms, if we assume an effective source plane velocity of 500 km s$^{-1}$, our simulated cadence approximately represents daily observations for our parameter range.
The full light curve would represent $\sim$4 years and does not include season gaps.
We acknowledge that this simulates a scenario not realizable by ground based follow-up, but note that we only focus on the ISCO crossing event which is a small portion of this light curve.

To produce caustic-crossing event light curves, we convolve our accretion disc surface brightness profile with the magnification map and take various trajectories across it.
Such example trajectories are shown in Fig.~\ref{Figure4} and a specific example of a simulated light curve is shown in Fig.~\ref{Figure5}.
We explore a wide range of impact angles and restrict ourselves to single caustic-crossing events.
Cusp crossings are omitted as these create different light curves than the more likely fold crossings (e.g. we use only paths above path B in Fig.~\ref{Figure4} and only those which cross one caustic fold).

\subsection{Light curve features}
\label{LCFeatures}
\begin{figure}
    \centering
    \includegraphics[width=0.47\textwidth]{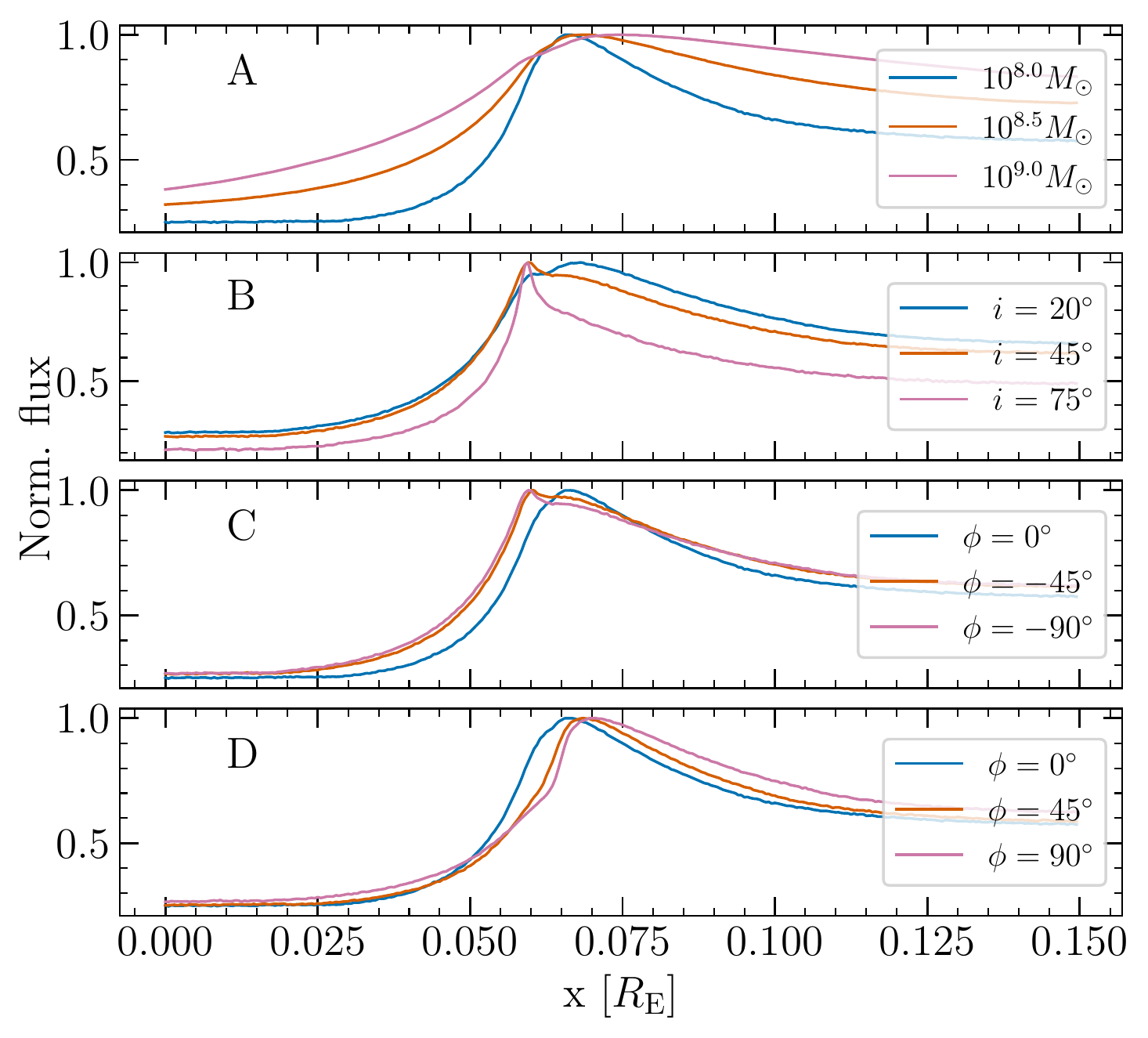}
    \caption{Effect of various parameters on light curves. Panel A) Light curves widen with $M_{\rm{BH}}$ for constant $i = 45\degree, \phi = 0\degree$. Panel B) Light curves are sensitive to disc asymmetry which increases with inclination. Each light curve has the approaching side (the brightened side) crossing first. Panel C) Effect of rotating the asymmetric accretion disc with respect to the caustic. In this panel, we rotate the disc such that the approaching side crosses the caustic first. Panel D) Same as panel C, but such that the receding side crosses the caustic first. These last two panels show that the double-peak feature may almost entirely vanish depending on parameter choices.}
    \label{Figure6}
\end{figure}

Fig.~\ref{Figure6} illustrates a selection of simulated caustic-crossing events spanning a sample range of explored parameters at $\lambda_{\rm obs}= 600 \text{nm}$, without photometric noise.
Redshifts were held constant at $z_{\rm{l}} = 0.5, z_{\rm{s}} = 2.0$ in all cases.
This choice led to ratios of $R_{\text{S}} / R_{\text{E}}=$ 0.006, 0.013, and 0.029 for log$_{10} (M_{\text{BH}} / M_{\odot}) = 8.0, 8.5, \text{ and } 9.0$, respectively.
A common feature observed in each light curve is a rise in the flux as the disc crosses over the caustic, peaking roughly when it crosses the ISCO region before dropping away.
The accretion disc size determines the steepness of the magnification change, such that smaller sources lead to steeper changes.
The most interesting part of these light curves is the fine structure of the signal that appears at the peak.
For a symmetric disc with a dark central region (e.g. a face on thin disc) the ISCO crossing results in a double-peak feature around the light curve peak.
The asymmetry in surface brightness produced by Doppler shifting, beaming, and local SMBH lensing may lead to the suppression of one of these peaks.
This is evident when comparing different orientations, as in the bottom two panels of Fig.~\ref{Figure6}.
Asymmetry increases between these peaks if the either the enhanced or suppressed side crosses the caustic first.

\begin{figure}
    \centering
    \includegraphics[width=0.47\textwidth]{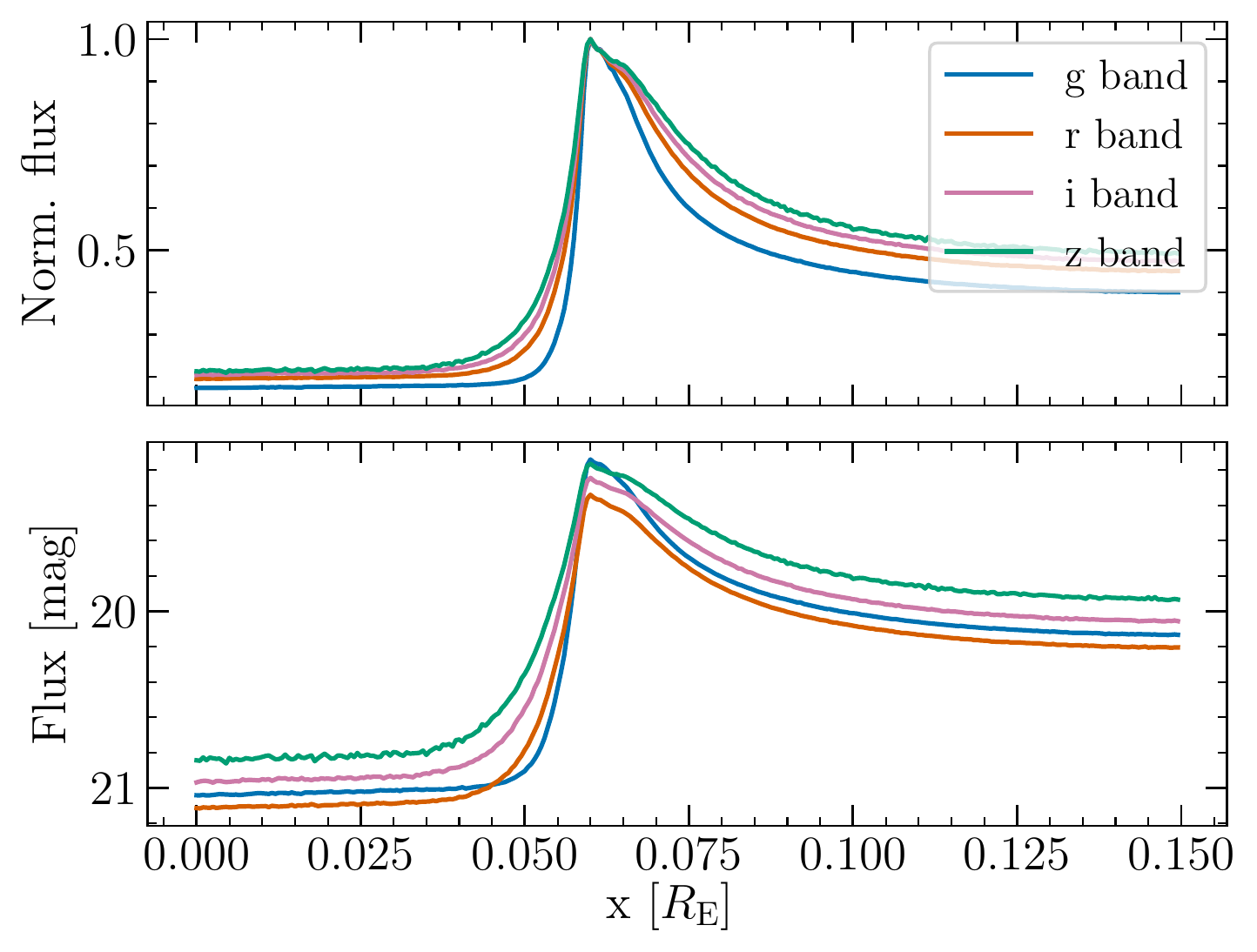}
    \caption{Mock light curves simulated for different optical bands for a caustic-crossing event with $M_{\rm BH} = 10^{8} M_{\odot}$, such as trajectory A in Fig.~\ref{Figure4} with the approaching side crossing first ($\phi = -90\degree$). Top: Normalized light curves show the general widening of the caustic-crossing event for the same ISCO size. Bottom: Flux of each simulated band in magnitude units. The zero point for each band was taken to be 3631 Jy, and conversions were calculated at wavelengths 477, 623, 762, 913 nm for each {\it g, r, i, z} band, respectively.}
    \label{Figure7}
\end{figure}

\begin{figure}
    \centering
    \includegraphics[width=0.47\textwidth]{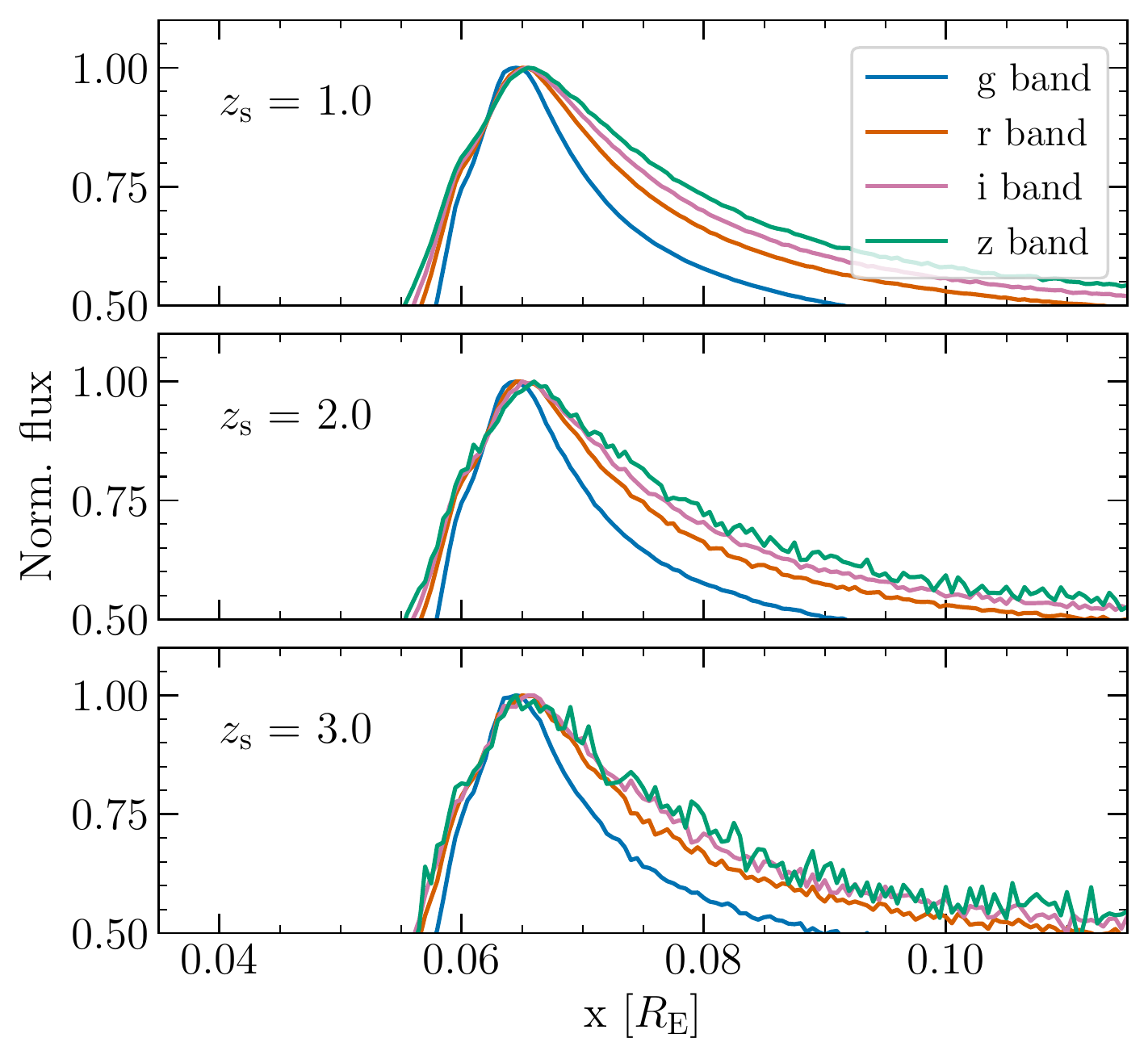}
    \caption{Peaks of a caustic-crossing event with photometric errors for discs at three different source redshifts. Each band is normalized independently to emphasise that the ISCO crossing feature remains the same, but the errors due to lower signal-to-noise ratios may obscure it. Each simulation used a $10^{8} M_{\odot}$ black hole approaching the caustic symmetrically ($\phi = 0\degree$) at $i = 45\degree$. At $z_{\text{s}}$ = 1.0, photometric noise is low in all bands. For $z_{\text{s}}$ = 2.0, the {\it g} and {\it r} bands appear smooth to the eye, while the {\it i} and {\it z} bands appear rough. At $z_{\text{s}}$ = 3.0, the effect of higher photometric noise is observed in all bands. The ISCO feature itself becomes less apparent with higher redshift.}
    \label{Figure8}
\end{figure}

Multi-wavelength observations can add significant information and potentially break a number of degeneracies.
Fig.~\ref{Figure7} demonstrates a single caustic-crossing event in various optical bands.
The width of the caustic-crossing event in the different simulated bands depends on the size of the accretion disc as viewed in those wavelengths, which has been used to measure the accretion disc temperature gradient \citep[e.g.][etc]{Wambsganss91, Anguita08, Bate18}.
However, the ISCO size is determined by the Kerr metric around the SMBH and is wavelength independent.
Fig.~\ref{Figure8} shows multi-wavelength light curves including our simulated noise model.
We highlight how with errors relative to 1 hour exposures, the ISCO crossing can potentially be resolved at shorter wavelengths even at high redshift ($z_{\rm s} \sim3$).

\section{Light curve analysis techniques}
\label{Analysissection}
The accretion disc/SMBH properties with the most impact on the fine structure of simulated light curves are the size of the ISCO and the flux asymmetry.
The ISCO and asymmetry manifests in the double-peak feature of the light curves as the distance and height difference between the two peaks (see Fig.~\ref{Figure6}).
The two peaks are separated by a reduction in flux due to the dark shadow of the black hole crossing over the caustic.
This separation is the full length of the projected ISCO in the direction that the accretion disc travels over the caustic.
We label this length as $L_{\rm{ISCO}}$ which has a maximum value of 12 $R_{\rm{g}}$.
Apart from these somewhat intuitive features, the combination of all model parameters (e.g. $M_{\rm BH}$, $i$, $\phi$) introduces complex, non-linear features in the light curves that are less straightforward to identify and interpret.

We approach the problem of measuring $L_{\rm{ISCO}}$ using two data analysis methods, which we call the second derivative method (Section \ref{Spline_Fit_Section}) and the wavelet method (Section \ref{sec:wavelet}).
These are straightforward methods that are easy to implement and interpret.
They work under the assumption that the ISCO is the only discontinuity in the accretion disc's brightness profile and is sufficiently reflected in the light curve.
We also use a machine learning approach on the entire non-linear parameter space in Section~\ref{machinelearning}.
Although more powerful, this approach requires a careful implementation and interpretation.

\subsection{Second derivative method} 
\label{Spline_Fit_Section}
We find the discontinuities associated with the ISCO crossing event can be located as a minimum in the second derivative of the continuous light curve due to the abrupt nature of the ISCO.
These minima correspond to the entry and exit of the ISCO across the caustic.
The separation of these minima is a measure of ISCO crossing time, or an ISCO size given an effective velocity.
In a sampled light curve, these can be detected by fitting a spline and taking the second derivative of this spline. 
We use splines because they are continuous, differentiable, and have the potential to reduce noise while still capturing the ISCO crossing inflection points.

\begin{figure*}
    \centering
    \includegraphics[width=0.97\textwidth]{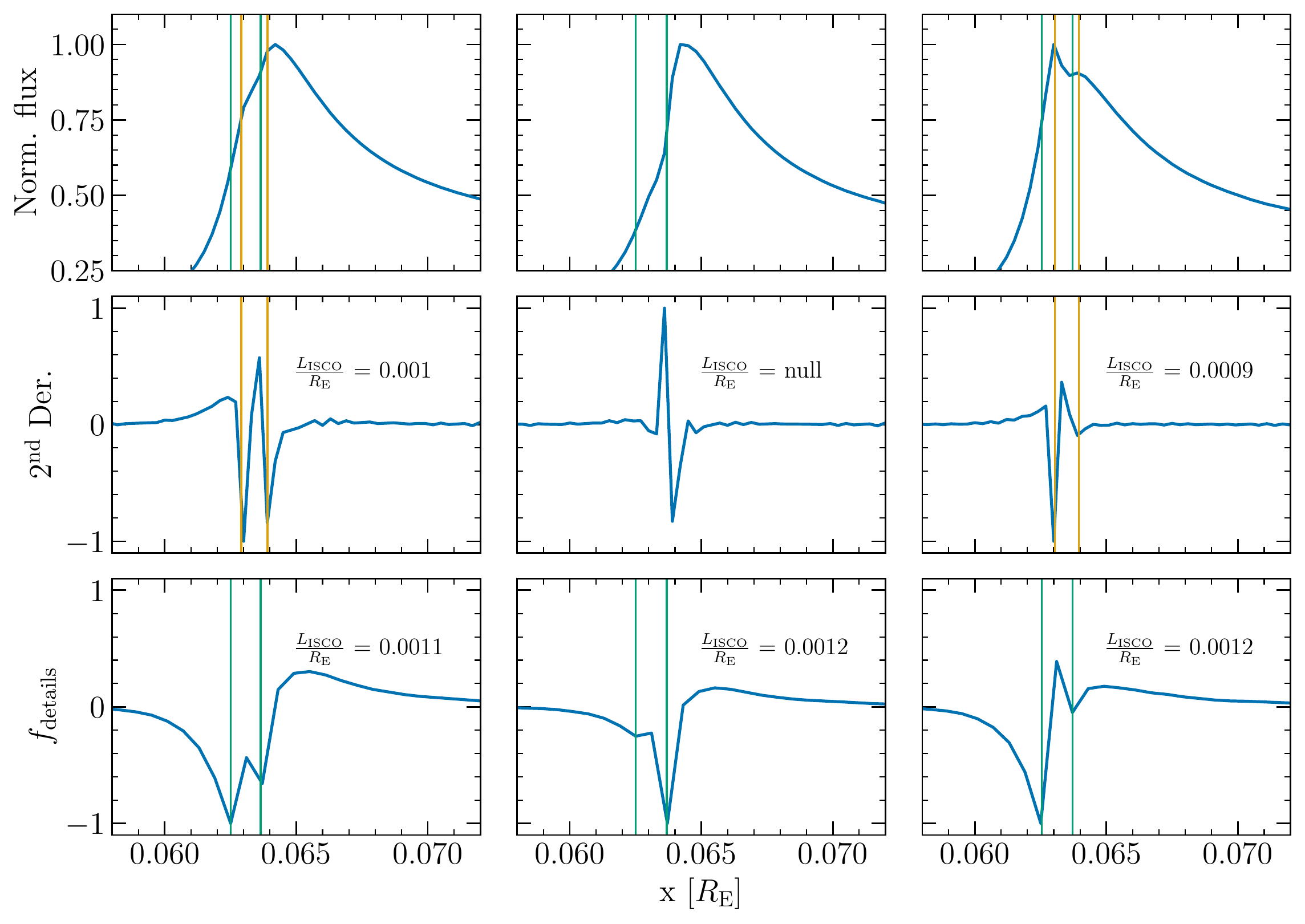}
    \caption{Measurements of $L_{\rm{ISCO}}$ for three sample light curves, each with $L_{\rm{ISCO}} = 0.00096 R_{\rm{E}}$. Top left: face on accretion disc. Top centre: moderately inclined accretion disc ($i = 45\degree$) with receding side crossing first ($\phi = +90\degree$). Top right: moderately inclined accretion disc ($i = 45\degree$) with approaching side crossing first ($\phi = -90\degree$). Centre row: measurements of $L_{\rm{ISCO}}$ using the second derivative method. The ISCO crossing signatures are emphasized with orange vertical bars extending to the light curves in the top row. Bottom row: measurements of $L_{\rm{ISCO}}$ using the wavelet method. The ISCO crossing signatures are drawn with green vertical bars. A slight offset is apparent when comparing these methods.}
    \label{Figure9}
\end{figure*}  
 
Our spline-fitting procedure utilizes \texttt{Scipy}\footnote{\url{https://scipy.org}} 's Univariate Spline.
\texttt{Scipy} is a open-source Python library designed to facilitate scientific computations~\citep{Scipy20}.
In fitting a spline to our simulated light curves, we take advantage of the smoothness condition governed by the parameter $s$ defined as:
\begin{equation}
    s \geq \sum_{i} (w_i (y_i - \hat{y}_i))^{2}
\end{equation}
where $w_i$ is the weighting of each point (taken to be uniform), $y_i$ is the true value, $\hat{y}_i$ is the spline fit value, and the sum is taken over all points in the light curve.
In general, a smaller value of $s$ means more knots are required in the spline to meet this constraint, while larger values allow less knots.
When the spline is constructed, the number of knots is increased and their positions are adjusted until this condition is met.
We note that this condition can always be met for any non-negative $s$, as the right hand side becomes zero for the case where all points are used as knots.

For a fit to be considered successful in measuring $L_{\rm{ISCO}}$, we assert that we must detect \emph{two distinct} minima in the second derivative of our spline. 
We define this as two local minima with amplitude greater than a predefined noise threshold normalized by the global minimum, and omit all cases with three or more local minima above this threshold.
Finding a successful measurement proceeds as such:

\begin{enumerate}[label=(\arabic*), itemindent=2em, leftmargin=0em, labelindent=0pt, labelwidth=!]
    \item The initial choice of $s$ is chosen arbitrarily large, such that a successful measurement will not occur.
    
    \item If there are too many/few detected minima in the spline, $s$ is decreased/increased randomly by 1 -- 10 per cent.
    
    \item The spline is fit to the data again with the new smoothness parameter and we check against our noise threshold condition. If this is met, the spline is returned as a successful measurement. Otherwise, we return to step 2.
    
    \item Steps 1-3 are repeated 100 times. The average measurement and standard deviation of $L_{\rm{ISCO}}$ is reported.
\end{enumerate}

We recognize there will likely be many values of $s$ for which our threshold condition is met. 
The impact this has on our measurement is minimized by the random adjustment of $s$ in step 2, and the repetition of this converging many times to different successful fits.
We also make note that if $s$ becomes too small (e.g. many knots are required to fulfil the smoothing condition), our threshold condition will not be met. 
Finally we note that while the simulated light curves have discrete values, the spline is continuous and measurements of $L_{\rm{ISCO}}$ are made without regard to the discrete simulated values.

The centre row of Fig.~\ref{Figure9} shows the second derivative of the spline fit to our simulated light curves of three caustic-crossing events after the fitting procedure has completed.
To make these measurements, we define the ISCO crossing length to be $L_{\rm{ISCO}}$ = $x_{2} - x_{1}$, where $x_{1}$ and $x_{2}$ are the positions of the second derivative minima for the spline marked with orange vertical lines.
Depicted are the three extreme cases of caustic-crossing events which have equal $L_{\rm{ISCO}}$ = 0.00096 $R_{\text{E}}$; the face on case, the receding side crossing first, and the approaching side crossing first.
The left most and right most splines successfully measure $L_{\rm{ISCO}}$ by meeting our noise threshold criteria chosen to be 5 per cent. 
The centre spline's second derivative did not meet the criteria for two distinct minima, as a third minimum was found below the noise threshold in the best fit.
Adjusting the smoothness parameter was found to suppress both minor minima in this case and the most prominent of the two could not be determined.

We note that in developing this method, we have assumed the fine structure will be detectable through the noise.
This method will not accurately measure the ISCO for every orientation, especially where one peak in the fine structure is suppressed (e.g. see the varying fine structures simulated in Fig.~\ref{Figure6}).
Furthermore, Fig.~\ref{Figure10} shows a region of parameter space where the ISCO crossing event may not be resolvable using these methods.

\subsection{Wavelet method}
\label{sec:wavelet}
The signature of the ISCO crossing is a small scale feature in these simulated light curves, with peaks lasting a fraction of an Einstein radius.
This allows it to be detected as a high frequency signal, potentially recoverable after applying a Fourier or wavelet transformation.
We utilize the latter as wavelet transforms retain spatial information unlike Fourier transforms. 
In making this transform, we project the light curve onto Daubechies wavelet basis vectors~\citep{Daubechies88} to detect the boundaries of the ISCO crossing.

An approximation of our light curve up to $N$ orthogonal basis vectors which probe increasingly higher frequencies may be described as the sum over projections onto these basis vectors~\citep{Mallat08}:
\begin{equation}
    f_{N} = \sum _{m = 0} ^{N-1} \langle f, g_{m} \rangle g_{m}
\end{equation}
where $f_{N}$ is the approximation of a function $f$ (e.g. our light curve) up to $N$ basis vectors, and $g_{m}$ are the basis vectors. 
If we let $N$ approach infinity, this sum converges with the initial function.
By truncating this sum at any point, the details of the initial function lost through making this approximation are then defined as:
\begin{equation}
    f_{\text{details}}(N) = \sum _{m = N} ^{\infty} \langle f, g_{m} \rangle g_{m} = f - f_{N}
\end{equation}
which may be interpreted as the residual between the true function and approximation.
Along our light curve, these wavelet coefficients are discretely calculated over a range of data points.
These coefficients are essentially the result of convolving a sliding kernel of basis vectors along the data, where we focus on recording the magnitude of $f_{\text{details}}(N)$.
We examine this high frequency information for a signal that corresponds to the discontinuities around $L_{\rm{ISCO}}$.
This is depicted in Fig.~\ref{Figure9} along the bottom row.
We define the detection of two distinct minima in this algorithm similarly to the second derivative method.
The method for determining the value of $N$ proceeds as follows:

\begin{enumerate}[label=(\arabic*), itemindent=2em, leftmargin=0em, labelindent=0pt, labelwidth=!]
    \item $N$ is set to one (e.g. our input data is projected onto a single basis vector which projects onto the lowest basis vector) and the details lost are calculated. We note that this should not give us a valid answer, but our goal is to find the smallest $N$ which satisfies our detection of distinct minima.
    \item $N$ is increased by one and the detail coefficients are calculated. This step is repeated until we detect two distinct minima in $f_{\text{details}}$.
\end{enumerate}

Unlike the second derivative method, projection onto these basis vectors does not require particular fitting so once the required number of basis vectors is met the calculation concludes.
We do not start at a large value of $N$ because at some scale the ISCO signature will be captured by the approximation light curve, and $f_{\rm{details}}$ will then only contain noise.
The module \texttt{PyWavelets}\footnote{\url{https://github.com/PyWavelets/pywt}}~\citep{Lee19} was used for this analysis, which is an open source python package designed for wavelet transformations.
As before, we make the assumption that there will be exactly two discontinuities arising from a single ISCO crossing event.

The bottom row of Fig.~\ref{Figure9} shows the detail coefficients lost once the distinct minima were located.
We measure $L_{\rm{ISCO}}$ = 0.0011 and 0.0012 $R_{\rm{E}}$, which is greater than expected.
We note that due to the discrete projections onto the wavelet basis vectors, the kernels have a finite size which will encounter the inflection sooner than expected.
This leads to minima preceding the actual ISCO crossing feature.

\begin{figure}
    \centering
    \includegraphics[width=0.45\textwidth]{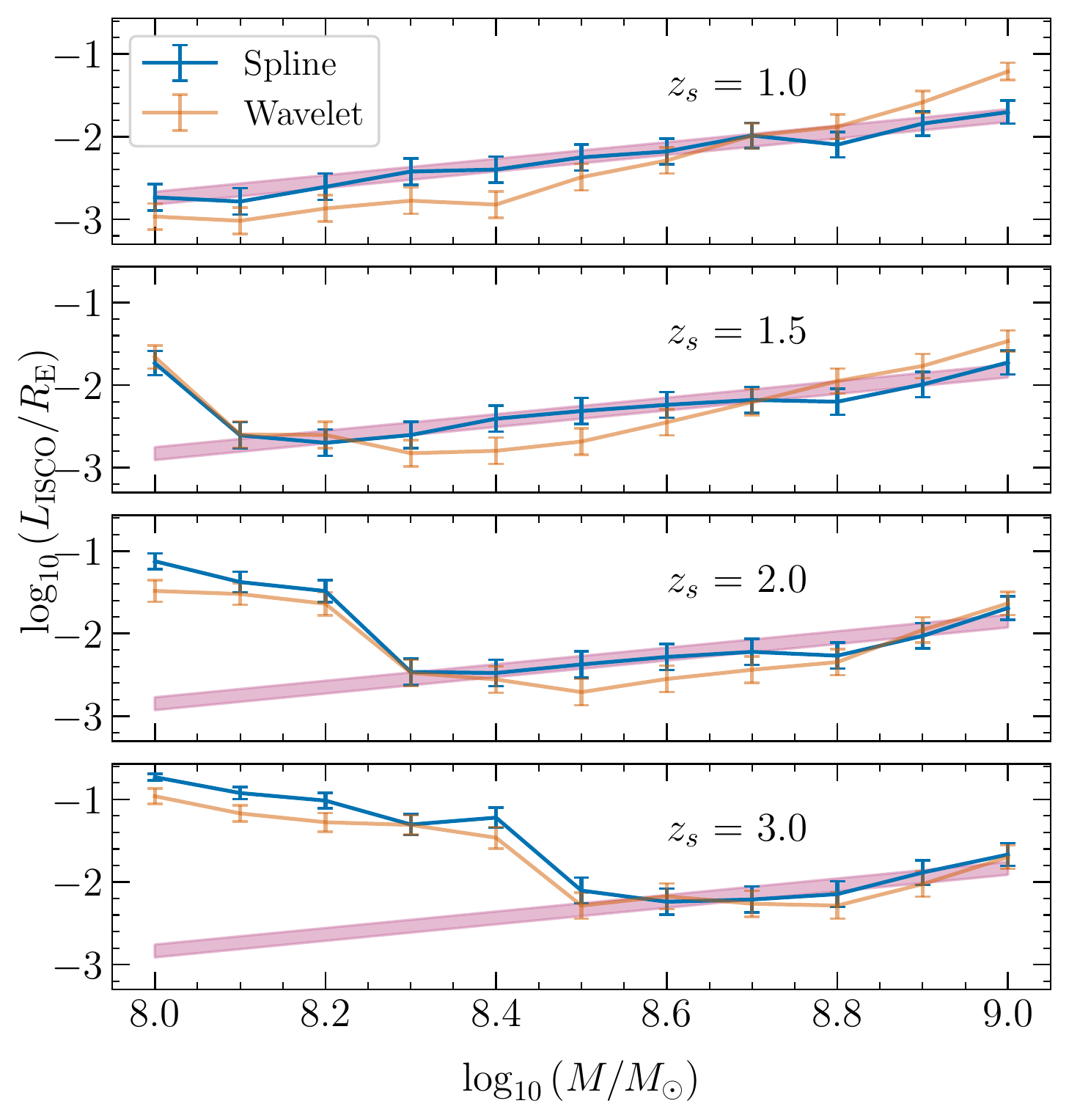}
    \caption{Average measured $L_{\rm{ISCO}}$ using the spline (blue) and wavelet (orange) methods as a function of black hole mass at various redshifts. Error bars are defined as 1 $\sigma$ in the measurements around the average measured ISCO. The shaded region represents the $L_{\rm{ISCO}}$ expected for each mass. Black holes of lower masses are not plotted here as they could not be measured accurately with these cadences and signal-to-noise ratios.}
    \label{Figure10}
\end{figure}

Similarly to the second derivative method, the wavelet method relies on the fine structure being detectable through the noise.
This assumption may break down when one peak in the fine structure is greatly suppressed due to orientation.
This also may break down for unfavourable SMBH mass and redshift combinations, as shown in Fig.~\ref{Figure10}.

\subsection{Machine learning method}
\label{machinelearning}
We use a regression approach to train a Convolutional Neural Network (CNN) on our simulated light curves.
While CNNs are traditionally used for image recognition~\citep{Lecun98, Lanusse18}, we take advantage of their ability to extract features even in one dimension.
We aim to directly predict four parameters: 1) projected ISCO crossing length $L_{\rm{ISCO}}$, 2) black hole mass $M_{\text{BH}}$, 3) inclination angle $i$, and 4) impact angle $\phi$.
The projected ISCO crossing length is precisely what was measured using the previous methods.
We then demonstrate machine learning has the ability to infer other parameters of the simulation.

When predicting these parameters, we face the issue that they are not measured on equivalent scales.
$L_{\rm{ISCO}}$ is predicted in $R_{\rm{E}}$, the mass is predicted in log space, and angles are predicted in degrees.
Machine learning works by minimizing a \emph{loss function} which quantifies a penalty based on far away a prediction is from the true value.
Creating these loss functions to cover various scales is non-trivial, so we avoided this issue by training multiple CNNs with identical architecture.
Each of these networks were trained to predict a single parameter in a series of \emph{targeted} samples.

The ranges of all parameters in each targeted sample remained the same as in Table~\ref{MLParams}.
However, the total number of light curves in each targeted data set was adjusted to accommodate higher resolution for angles $i$ and $\phi$ (e.g. the strides $\Delta i$ and $\Delta \phi$ were adjusted).
This allowed us to: 1) Resolve these angles at higher resolution when targeted and 2) Increase the learning rate when these parameters were not targeted.
Our choices omit the $i = 0\degree$ case, where all $\phi$ values become impossible to distinguish.
We include the $\phi = 0\degree$ case when taking either stride through the parameter space.

Each light curve is introduced to the CNN as an array of fixed length with values corresponding to the normalized amplitude of the light curve.
This was done because a CNN requires equal step sizes due to the discrete convolutional kernels used.
Normalization was done as a pre-processing step such that each light curve contains values between zero and one.
This is standard practice in machine learning and helped our predictions converge across all validation sets.
However, normalization can potentially effect how the machine sees the asymmetry of the ISCO crossing event.
We found that this loss of this information does not induce any particular biases with the predictions of our targeted parameters.

Each set of light curves contained a total of $\sim$ 100,000 to $\sim$ 400,000 light curves, depending on which parameter was targeted.
A random subset of 20 per cent of the light curves were set aside and strictly used for validation, while the remaining were used in training.
Splitting the data set like this allows us to determine how well the network learned by applying it to light curves not previously trained on.

We now outline our network architecture which is summarized in Table~\ref{CNNArch}.
Our input is first passed into three stacked convolutional and pooling layers.
These convolutional layers use shift invariant kernels that extract features from the light curves, where a total of 32 kernels were used leading to an increase of dimensionality in the outputs.
The output of the convolution is passed through a rectified linear unit (ReLU) activation function~\citep{RELU}, which introduces non-linearity into the network by setting all negative values to zero.
The max pooling layers then act to reduce the size of the output in order to probe progressively larger size scales using the same kernels.
Both convolutional and max pooling layers include zero padding which assured the outputs of these layers remained consistent in size.
After the convolutional and max pooling layers, the outputs are flattened into one dimension then passed through a sequence of three fully connected layers (dense layers).
There is a final fully connected layer with a single output that represents the predicted value of the target parameter.
This target parameter can represent the mass, inclination, impact angle, or ISCO length.
We tested multiple other CNN architectures and found no significant change in performance, so a simple and efficient one was chosen.

\begin{table}
    \centering
    \caption{Convolutional Neural Network architecture.}
    \begin{tabular}{l c c}
     Type & Output shape & Parameters \\
     \hline 
     Input & 150 & -\\[0.25ex] 
     Convolution & (150,32) & 128 \\ [0.25ex] 
     ReLU & - & - \\[0.25ex] 
     MaxPooling & (75,32) & - \\[0.25ex] 
     Convolution & (75,32) & 3,104 \\[0.25ex] 
     ReLU & - & - \\[0.25ex] 
     MaxPooling & (38,32) & - \\[0.25ex] 
     Convolution & (19,32) & 3,104 \\[0.25ex] 
     ReLU & - & - \\[0.25ex] 
     MaxPooling & (19,32) & - \\[0.25ex] 
     Flatten & 608 & - \\[0.25ex] 
     Fully Connected & 64 & 38,976 \\[0.25ex] 
     ReLU & - & - \\[0.25ex] 
     Fully Connected & 32 & 2,080 \\[0.25ex] 
     ReLU & - & - \\[0.25ex] 
     Fully Connected & 32 & 1,056 \\[0.25ex] 
     ReLU & - & - \\[0.25ex] 
     Fully Connected & 1 & 33 \\[0.25ex] 
     Total &  & 48,481 \\[0.25ex] 
    \end{tabular}
    \label{CNNArch}
\end{table}

To train our CNN, we use a mean squared error loss function, defined as:
\begin{equation}
    \mathcal{L}(y,\hat{y}) = \frac{1}{N}\sum_{i=1}^N (y_i-\hat{y}_i)^2 , 
    \label{MSE}
\end{equation}
where $y$ is the true value (the training label), $\hat{y}$ is the value predicted by the network, and $N$ is the size of the sample.
This loss function is minimized using an Adam optimizer~\citep{kingma2017adam}, a type of stochastic gradient descent.
Using this optimizer introduces a few hyperparameters such as the learning rate and batch size.
The learning rate acts as a coefficient which helps determine how much any individual weight can change at a step in the training process.
If this is too high, the weights will update too quickly and they will not converge to an optimal value.
On the other hand, if it is too low the weights will take too long to converge or may get stuck in a local minima--never to find the true minimum.
The batch size determines how many light curves are passed through the network in the training step before internal weights are updated.
We chose to use a constant learning rate of 0.001 and a batch size of 32, which allows learning to occur at a reasonable pace without greatly overshooting the optimal values.
We trained the network over the entire sample for 200 iterations known as epochs to assure the validation loss had converged.

\begin{table}
    \centering
    \caption{Parameters used in creation of simulated light curves, all ranges are inclusive. Multiple strides are taken through $i$ and $\phi$ data ranges depending on data set (see text for details).}    
    \begin{tabular}{lcc}
    Parameters & Range & Stride \\
    \hline
    $\text{log}_{10}(L_{\rm{ISCO}} / R_{\rm E})$ & $-3 , -1.5$ & -- \\
    $\text{log}_{10} (M_{\rm BH} / M_{\odot})$ & $7.7 , 9.7$ & 0.1 \\
    $i$ & $10\degree , 80\degree$ & $5\degree$ / $10\degree$ \\
    $\phi$ & $-90\degree , +90\degree$ & $5\degree$ / $20\degree$ \\
    $z_{\rm s}$ & 1 , 3 & 0.5\\
    \end{tabular}
    \label{MLParams}
\end{table}

\begin{figure*}
    \centering
    \includegraphics[scale=0.91]{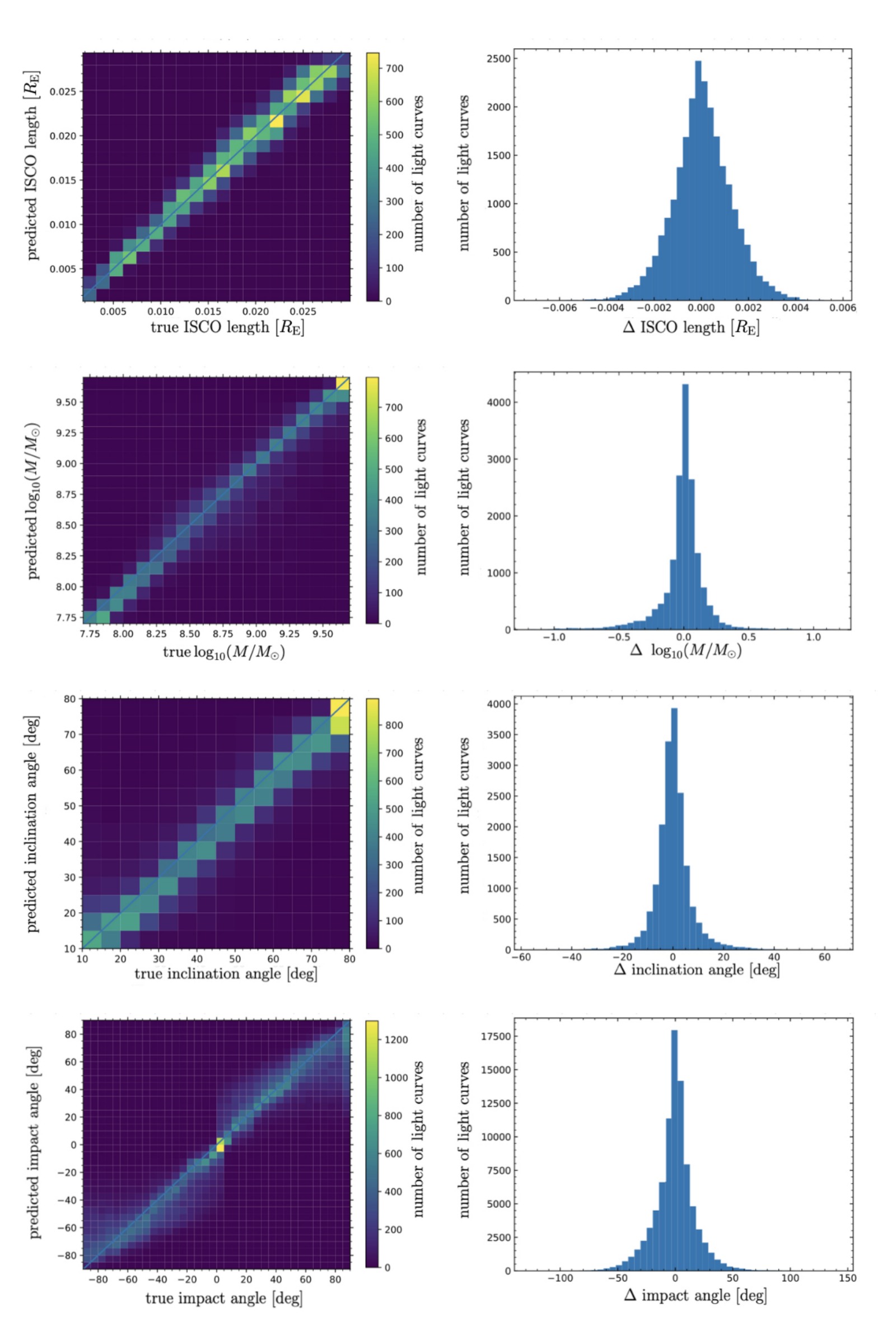}
    \caption{Left: Predictions vs truths for each targeted parameter. Right: Histograms of the differences between predicted and true values. Means and standard deviations are presented in Table \ref{uncertainty_table}.}
    \label{Figure11}
\end{figure*}

\begin{figure*}
    \centering
    \includegraphics[width=0.93\textwidth]{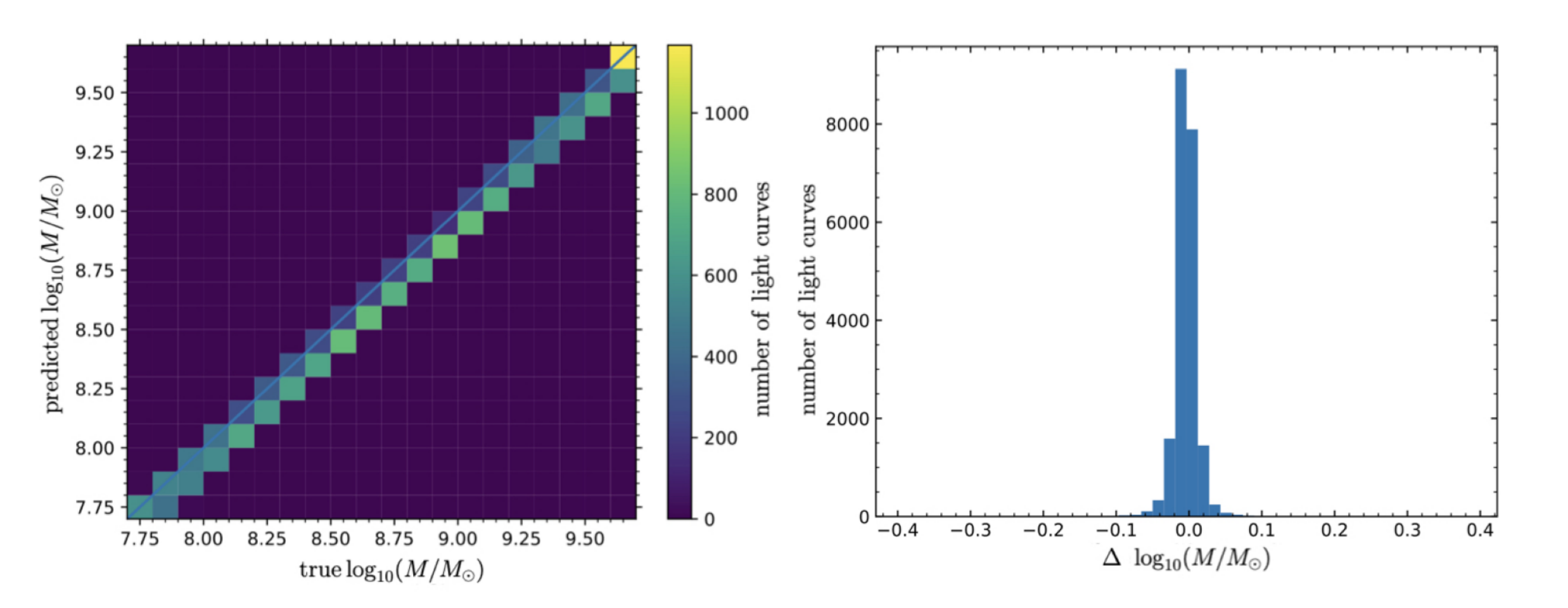}
    \caption{Predictions vs truths and histogram for the prediction of the mass label over the baseline set. The bias and uncertainty are given in Table~\ref{test_set_uncertainty_table}.}
    \label{Figure12}
\end{figure*}

\begin{figure*}
    \centering
    \includegraphics[width=0.93\textwidth]{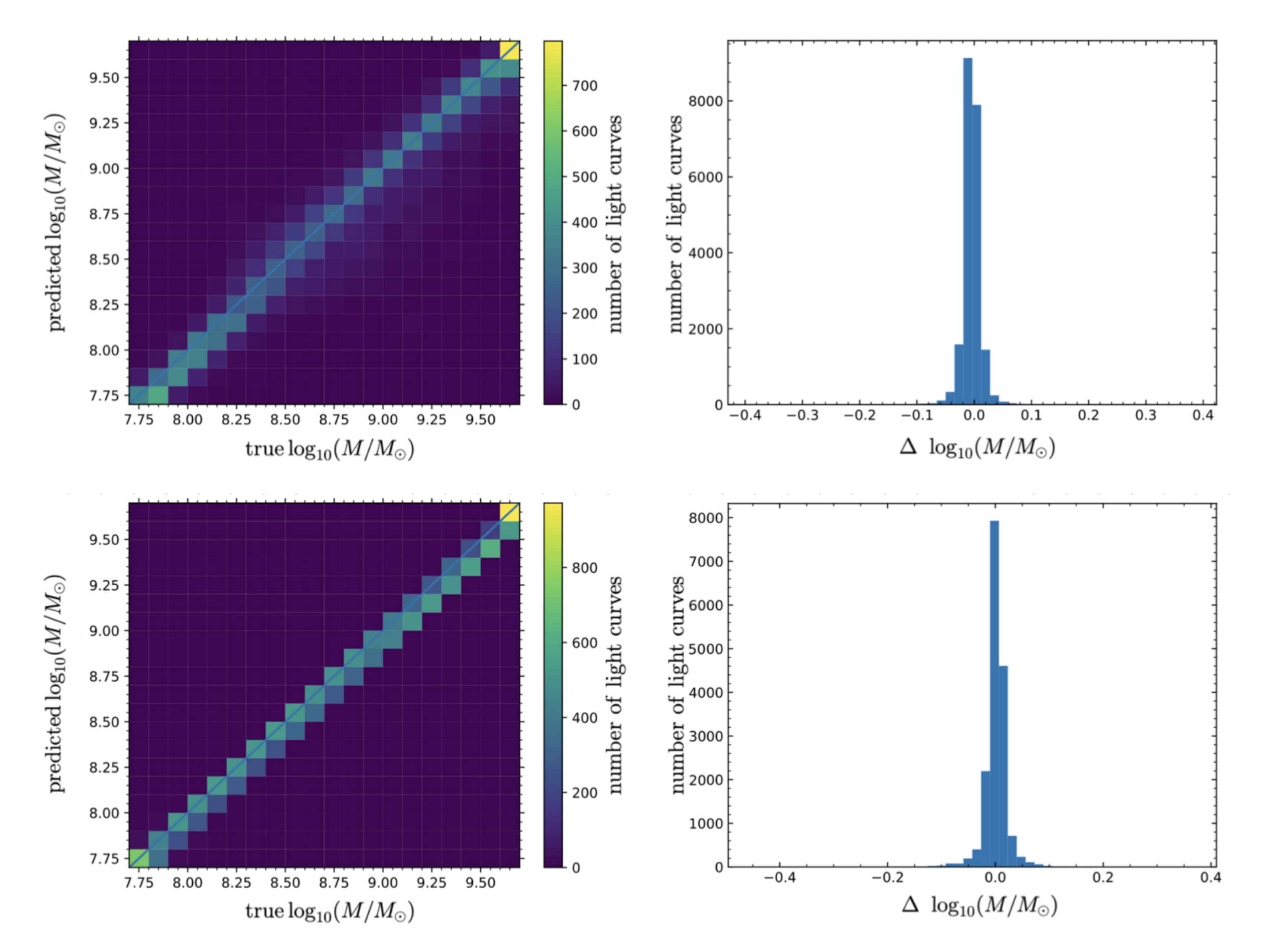}
    \caption{Predictions vs truths and histograms for the alternative sets, which included noise levels based on best case Gemini-like observations. Top row: Predictions on alternative set 1 where $R_{\rm{S}}$ was held constant. Bottom row: Predictions on alternative set 2 where $R_{\rm{in}}$ was held constant. The biases and uncertainties are given in Table~\ref{test_set_uncertainty_table}.}
    \label{Figure13}
\end{figure*}

To understand what features the network learns from, we tested our network's ability through a pair of \emph{alternative simulations}.
The goal of this testing is to gain insight to what features of the light curve are most important for inferring $M_{\rm{BH}}$.
In our first alternative simulation, the accretion disc size $R_{\rm{S}}$ was held constant to that of a black hole with mass $10^{8} M_{\odot}$, while the labeled mass corresponds to the artificially adjusted ISCO size. 
This probes the ability for our network to extract small scale features at the size of the ISCO.
Predictions on the second alternative simulation aimed to test the reverse--the ISCO was artificially held constant to that of a black hole with mass $10^{8} M_{\odot}$, while the labeled mass defines $R_{\rm{S}}$.
This second set tests if the CNN is predicting based on large scale features such as the apparent accretion disc size.
To make these adjustments, $R_{\rm{in}}$ was no longer defined as $6 GM_{\rm{BH}}/c^{2}$, but chosen independently of $M_{\rm{BH}}$.
We note that by artificially modifying these accretion discs, the value of $R_{\rm{S}}$ is slightly changed.
However, the inner accretion disc contributes a very small amount of flux at optical wavelengths and we normalize all light curves.
These modifications should not introduce undesirable artefacts or biases.
Our modified accretion disc images are used to prepare a full set of microlensing light curves spanning the parameter space, and our noise model is applied to the simulated flux.
As before, our CNN is then trained on 80 per cent of these light curves and we test on the remaining 20 per cent.
The results of these tests are presented in Section~\ref{CNNTestSetResults}.

We also tested the ability of our network to extend predictions down to significantly lower signal-to-noise ratios as outlined in Appendix~\ref{IncreasedNoiseAppendix}.
The goal of this was to demonstrate that the CNN is not as sensitive to the signal-to-noise level as our previous two methods.
After demonstrating this, we increased the noise levels of the light curves shown to the CNN in order to generalize this method to more realistic data.

\section{Results}
\label{resultssection}

The results of this study on simulated light curves are organized in the following manner.
Section~\ref{ISCOlengthresults} contains the results of measuring the $L_{\rm{ISCO}}$ using all available methods.
When the ISCO length is measured using the second derivative and wavelet methods, the data are separated by redshift to emphasize regimes where the methods are accurate.
In the machine learning case we do not make this separation when predicting the ISCO as there is no clear distinction where the network fails.
Section~\ref{OtherResults} contains the predictions of the CNN on the black hole mass, inclination angle, and impact angle.
Section~\ref{CNNTestSetResults} describes the results of the CNN's mass inference from our alternative sets, and compares these results with baseline results.

\subsection{Measurement of projected ISCO length}
\label{ISCOlengthresults}

Fig.~\ref{Figure10} presents the ISCO crossing length measured using the second derivative and wavelet methods for four source redshifts (1.0, 1.5, 2.0, and 3.0) as a function of $M_{\rm{BH}}$.
The error bars represent the standard deviation around mean measurements where the average standard deviation is 0.27 dex for the second derivative method, and 0.32 dex for the wavelet method.
The shaded region represents the range of true ISCO lengths as a function of mass at each source redshift for all geometric orientations considered.

The second derivative method is successful at determining the projected ISCO length for a wide range of configurations in parameter space.
At $z_{\mathrm{s}}=1.0$, this method is successful at recovering $L_{\rm{ISCO}}$ for $M_{\rm{BH}} \geq 10^{8.0} M_{\odot}$.
For $z_{\mathrm{s}}=1.5$, we find that this method is successful for $M_{\rm{BH}} \geq 10^{8.1} M_{\odot}$.
At source redshifts 2.0 and 3.0, this lower limit increases to masses of $10^{8.3} \text{ and } 10^{8.5} M_{\odot}$, respectively.
The lower bound of measurable $L_{\rm{ISCO}}$ is due to the simulated noise levels, which increases with redshift and decreases with SMBH mass.
As such, this could potentially be resolved with higher signal-to-noise observations, but would not be likely to be realized by ground based follow-up due to our pristine follow-up conditions simulated.
We note that when measuring the projected ISCO length from light curves using this method, there are some events where our algorithm does not find a successful fit.
The most common orientation which leads to the failure of this ISCO measurement is when one peak is greatly suppressed due to moderate/high $i$ and $\phi \sim \pm 90\degree$.

The results of the wavelet method are presented in Fig.~\ref{Figure10} as well.
This method has potential to effectively measure the ISCO length imprinted on light curves for many configurations.
It suffers from the same issue of not being able to measure the low $M_{\text{BH}}$, high $z_{\text{s}}$ region of parameter space as in the second derivative method.
The average uncertainty for the wavelet method was found to be 0.32 dex, which was greater than that of the second derivative method.
Both methods have the same lower limit to where they lose the ability to measure $L_{\rm{ISCO}}$.

The top row of Fig.~\ref{Figure11} shows the predictions of $L_{\rm{ISCO}}$ made by the CNN. 
Since the CNN has the ability to predict $L_{\rm{ISCO}}$ throughout the parameter space, we did not separate the training sets into multiple subsets.
In testing the CNN, we did not have the same regions of parameter space where $L_{\rm{ISCO}}$ was immeasurable as in the previous methods.
Table~\ref{uncertainty_table} reports the mean offset and the standard deviation for the predicted $L_{\rm{ISCO}}$.
The bias in our predictions of $L_{\rm{ISCO}}$ is found to be insignificant, and the uncertainty is found to be on the order of our spacing between data points.
We find that these are significantly better than those measurements from the second derivative method and the wavelet method, even when we include much higher photometric error as described in Appendix~\ref{IncreasedNoiseAppendix}.

In comparing the results from each of these methods, we find that the second derivative method led to an unbiased result for signal-to-noise ratios high enough to allow for the direct measurement of $L_{\rm{ISCO}}$. 
The wavelet method was also found to measure $L_{\rm{ISCO}}$ in the same parameter space as the second derivative method, albeit with higher recorded uncertainties.
Our CNN was found to predict unbiased results with low uncertainty in all cases.

\subsection{Mass, inclination, and impact angle predictions by the CNN}
\label{OtherResults}

Machine learning has the ability to predict beyond $L_{\rm{ISCO}}$ due to the non-linear connections drawn between various data points.
In addition to the ISCO length, we trained neural networks to predict the black hole mass, the inclination angle, and the impact angle.
Each of these parameters jointly determine the value of $L_{\rm{ISCO}}$ in the source plane, but are not explicitly reflected in light curves.
Each network had the same architecture and was trained independently over a full targeted sample of 100,000 to 400,000 light curves described in Section~\ref{machinelearning}.
The results of these predictions are presented in Fig.~\ref{Figure11} and the bias and uncertainty of each measured parameter are given in Table~\ref{uncertainty_table}.
In all cases the predictions and truths are diagonal, meaning the networks did not suffer major systematic biases and were able to accurately predict the value of the targeted parameter.

The uncertainties of our predictions give us a measure of how confident the system is at predicting any variable.
We found that the CNN can constrain the mass to $\pm$ 0.185 dex (1 $\sigma$) for these simulations.
For inclination, we find the uncertainty to be $\pm 7.39\degree$.
The impact angle was found to have uncertainty of $\pm 17.9\degree$.
Our mean predictions were not found to deviate significantly from the true parameter values in any case.

As the CNN learns to make predictions only within the range of our prior, we see excess predictions near the edge of the parameter space where the prior is truncated for the mass and inclination angle (e.g. Fig.~\ref{Figure11}).

\subsection{Inference of mass from alternative simulations}
\label{CNNTestSetResults}

In order to explore which light curve features our CNN extracts parameters from, we attempted to predict the mass from two alternative simulations.
Mass was chosen as the target label as it determines the size scale of the accretion disc in the source plane.
In each of these alternative simulations, we adjusted either the small scale structure (e.g. the ISCO size) or the large scale structure (e.g. the accretion disc size) while holding the other constant.
We compare the results of these predictions to a baseline where everything is simulated as described in Section~\ref{LCSubSection} (e.g. without the additional photometric uncertainty).
The predicted vs true values of our baseline are shown in Fig.~\ref{Figure12}, where our prediction of the mass label was marginally biased toward lower values by -0.004 dex, and the uncertainty was found to be $\pm$ 0.017 dex.
The results of this experiment are summarized in Table~\ref{test_set_uncertainty_table}.

In the first alternative simulation, each accretion disc size was held constant while the mass label only effected the ISCO size.
The results of predicting the mass label from this alternative set are shown in the top row of Fig.~\ref{Figure13}, where the bias was found to be -0.001 dex and the uncertainty was found to be $\pm$ 0.067 dex. 
When we compared this to the baseline mass predictions, the bias remained insignificant.
This increase in uncertainty with respect to the baseline is expected as the only information pertaining to the mass label is encoded in the ISCO crossing event.
Through this test, we show that this CNN can accurately predict our mass label based \emph{only on small scale features} in the light curve.

With the second alternative simulation, the ISCO size was held constant while the mass label effected the accretion disc profile.
The results of these predictions are given in the bottom row of Fig.~\ref{Figure13}.
The bias was found to be +0.001 dex, and the uncertainty was $\pm$ 0.026 dex.
As in the other cases, the results are not significantly biased.
We find the uncertainty to be greater than the baseline but less than the first alternative simulation.
We conclude that the CNN has an easier time predicting $M_{\rm{BH}}$ from large scale features rather than small scale features, but \emph{both} scales are used when the network makes its predictions.

\begin{table}
\centering
 \caption{Bias and $1\sigma$ uncertainties across the validation set in each networks' parameter inference. The related plots comparing predicted vs true values are presented in Fig.~\ref{Figure11}.}
 \begin{tabular}{l l l l} 
 Parameter & Bias & Uncertainty \\ [0.25ex]
 \hline
 $L_{\rm{ISCO}}$ [$R_{\rm{E}}$] & +0.00004 & 0.00126   \\ [0.35ex] 
 $\log_{10}(M_{\rm{BH}}/{M_{\odot}})$ & -0.006 & 0.185 \\[0.35ex] 
 Inclination angle $i$ & +0.14$\degree$ & 7.39$\degree$ \\[0.35ex] 
 Impact angle $\phi$ & -0.58$\degree$ & 17.9$\degree$ \\[0.35ex] 
\end{tabular}
\label{uncertainty_table}
\end{table}

\begin{table}
\centering
 \caption{Bias and $1\sigma$ uncertainties across the validation set for the baseline and alternative simulations when predicting $\log_{10}(M_{\rm{BH}}/{M_{\odot}})$. All units are dex. The related plots comparing predicted vs true values are in Figs.~\ref{Figure12} and ~\ref{Figure13}.}
 \begin{tabular}{l l l l} 
 Simulation set & Bias & Uncertainty \\ [0.25ex]
 \hline
 Baseline set & -0.004 & 0.017  \\ [0.35ex] 
 Alternative set 1 & -0.001 & 0.067 \\[0.35ex] 
 Alternative set 2 & +0.001 & 0.026 \\[0.35ex] 
\end{tabular}
\label{test_set_uncertainty_table}
\end{table}

\section{Applications to QSO 2237+0305}
\label{Results2237}

Multiple high magnification events of QSO 2237+0305 have been observed, three of which are accepted to be caustic-crossing events.
These events were observed by the \href{https://ogle.astrouw.edu.pl}{OGLE} ~\citep{Wozniak00, Udalski06} and the \href{https://vela.astro.uliege.be/themes/extragal/gravlens/bibdat/engl/lc_2237.html}{GLITP} collaborations~\citep{Alcalde02}.
However, the cadence of each individual event is insufficient for parameter inference using our methods which require the high cadence follow-up of a caustic-crossing event.
To circumvent this issue, we instead analyze the composite light curve synthesized by ~\citet{Mediavilla15b}.

The composite light curve of ~\citet{Mediavilla15b} was produced in the following manner: each microlensing event was defined as the flux difference between two images.
For each event, one image was assumed to experience a caustic-crossing event while the other image experiences a slow variation in flux known as the \emph{microlensing baseline}.
This microlensing baseline was approximated to be linear over the event, and was subtracted.
The light curves were normalized in amplitude, regularly binned in 5 day intervals, and the average of each bin was taken.
The estimates on error were defined as the average of the difference between adjacent points separated by 2 days or less.
The final composite light curve represents a caustic-crossing event built from these three events with a conservative error estimate.

Since we are analyzing combined data, we must consider potential issues which may arise.
Caustic features are rarely aligned with each other except for the case of high external shear.
The duration of the microlensing event depends on the perpendicular component of velocity with respect to the caustic, $v_{\rm{perp}}$~\citep[e.g.][]{Poindexter10b}.
The fact that the motion of QSO 2237+0305 in the source plane is not well constrained enhances this potential issue.
However, as shown in figure 2 of~\citet{Mediavilla15b}, the caustic-crossing events we focus on each appear to last $\sim$ 100 days.
We take this as evidence that the accretion disc of QSO 2237+0305 does not travel parallel to the relevant caustic features, and all light curves have a similar $v_{\rm{perp}}$.
The difference in $v_{\rm{perp}}$ between events may differ by as much as $\sim$ 15 per cent and is small in comparison to our effective velocity range derived below in Equation (\ref{v_effective}).
This additional uncertainty will marginally effect our uncertainty in the mass estimation of the SMBH in QSO 2237+0305.

The analysis methods presented in Section~\ref{Analysissection} have been acting on simulated signals that are normalized by the Einstein radius.
To address the microlensing degeneracies described in Section~\ref{LCSubSection} and bring our simulations to the context of these observations, we must assume a microlens mass and have an effective velocity model.
We choose $M_{\rm{l}}$ to be 1 $M_{\odot}$ to remain consistent with our simulations and point out this is greater than typical estimates~\citep{Kochanek04, Anguita08, Eigenbrod08, Poindexter10a}. 
Along with our assumed cosmology and knowledge of the redshift configuration, this leads to $R_{\rm{E}} = 1.8 \times 10^{15}$ m.

We use the velocity model from~\citet{Neira20} to estimate the effective velocity $v_{\rm{eff}}$ in the source plane.
This takes into account the CMB dipole, the bulk velocity of compact objects within the lensing galaxy, and the peculiar velocity of the lens and source~\citep{Kayser86, Wyithe00a}.
The effective velocity is calculated as:
\begin{equation}
    \label{v_effective}
    v_{\rm{eff}} = \frac{v_0}{1+z_{l}} \frac{D_{ls}}{D_{ol}} - \frac{v_{\star}}{1+z_{l}} \frac{D_{os}}{D_{ol}} + v_{g} ,
\end{equation}
where $v_{0}$ depends on the CMB dipole, $v_{\star}$ depends on the microlens velocity dispersion, and $v_{g}$ depends on the peculiar velocities for the lens and source.
The value of $v_{0}$ = 58 km s$^{-1}$ is used to represent the contribution from the CMB dipole at the on-sky location of QSO 2237+0305~\citep{Kogut93}.
To build up an effective velocity distribution, we draw values of ${v_{\star}}$ and $v_{g}$ from normal distributions. 
These distributions are centred at zero and have widths of $\sigma_{\star}$ = 200 and $\sigma_{g}$ = 3023 km s$^{-1}$ respectively.
We note that Equation (\ref{v_effective}) is a vector sum, so we include uniformly distributed random directions with respect to the CMB dipole for each velocity.
We draw 10,000 realizations of $v_{\rm{eff}}$ to build our velocity distribution.
Finally, we discard the direction of $v_{\rm{eff}}$ because we only focus on the case of a single caustic-crossing event.
Our final distribution of $v_{\rm{eff}}$ is found to be centred at 3357 km s$^{-1}$, with a 1 $\sigma$ standard deviation of 2088 km s$^{-1}$.

One aspect we have neglected in our simulations is the effect of the black hole's spin $a_{*}$ on the ISCO.
A relatively high spin admits smaller prograde orbits around the black hole, reducing the size of the ISCO.
We choose to use spin information following the results of~\citet{Reynolds14b}, who measure $a_{*}$ = $0.74^{+0.06}_{-0.03}$ using relativistic X-ray disc reflection models and archival \emph{Chandra} observations.
Based on this spin estimate, QSO 2237+0305 has $L_{\rm{ISCO}} \leq 6.41^{+0.28}_{-0.60} R_{\rm{g}}$, where the inequality allows for the projected ISCO compression due to orientation effects.
This significantly differs from the Schwarzschild case where $L_{\rm{ISCO}} \leq 12 R_{\rm{g}}$.
Taking this into consideration effects our derived black hole mass with each method.

Another aspect which impacts the observed $L_{\rm{ISCO}}$ is the inclination of the disc.
As shown in Fig.~\ref{Figure2}, the greatest compression of the ISCO's minor axis may only be $\sim$ 0.7 for highly inclined discs due to light bending effects. 
It is believed QSO 2237+0305 is viewed relatively face-on, and we have no information regarding the impact angle $\phi$ which likely differs between the caustic-crossing events.
We do not assume $L_{\rm{ISCO}}$ is significantly reduced from orientation effects after averaging, and assert $L_{\rm{ISCO}} = 6.41^{+0.28}_{-0.60} R_{\rm{g}}$.

\begin{figure}
    \centering
    \includegraphics[scale=0.5]{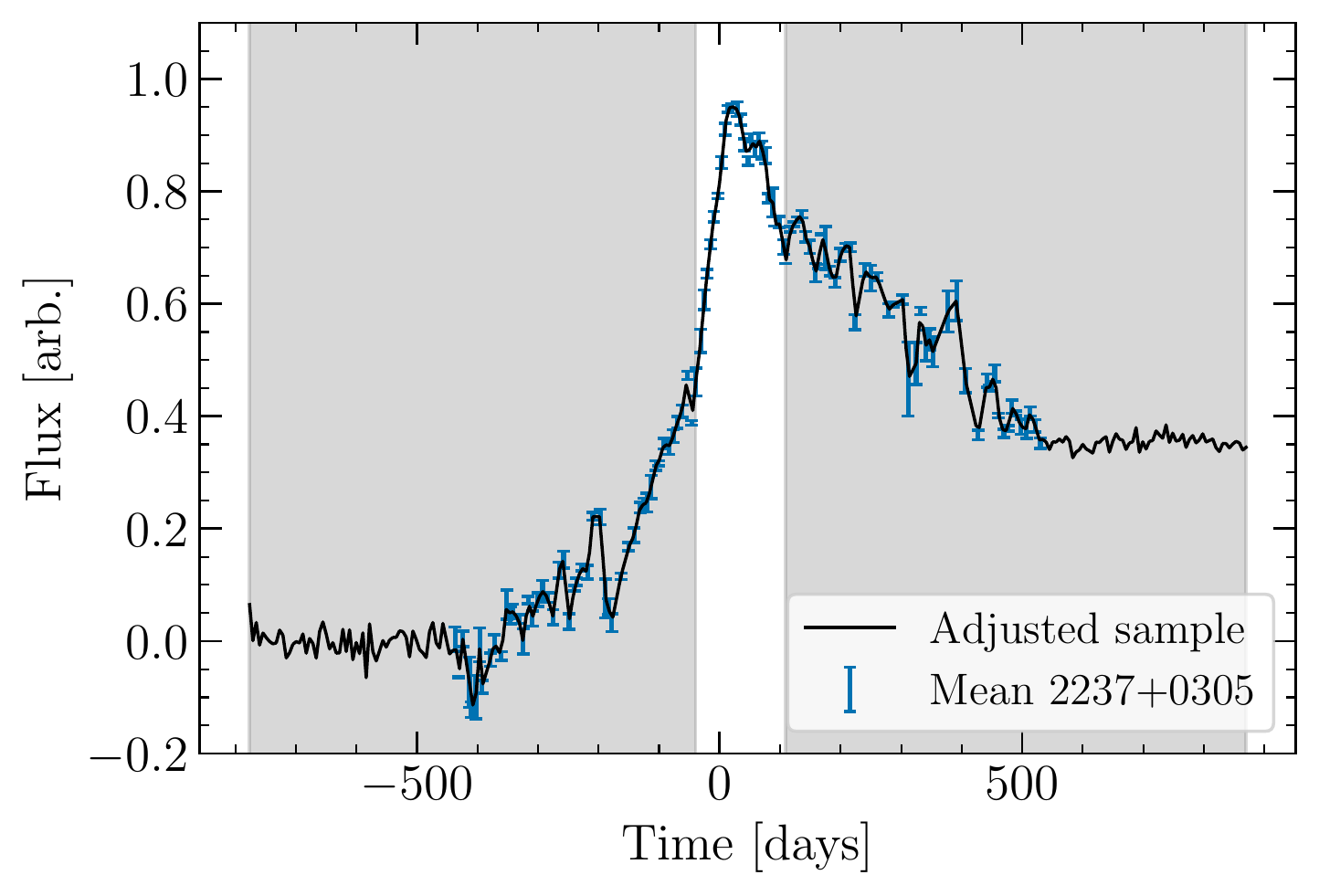}
    \caption{Representative caustic-crossing event of QSO 2237+0305. The grey shaded region represents the area expected to be unaffected by ISCO region and ignored for the purposes of the second derivative and wavelet analysis. The black curve is seen by the CNN and is represented by an array of 150 evenly spaced points. The representative light curves shown to the CNN will cover ($944^{+1554} _{-362}$) days based on $v_{\rm{eff}}$.}
    \label{Figure14}
\end{figure}

To measure the ISCO crossing event using the direct methods described in Section~\ref{Analysissection}, we assert the ISCO only imprints itself near the peak of the caustic-crossing event as seen in simulations.
For the second derivative and wavelet methods we exclude the grey shaded regions of Fig.~\ref{Figure14} and instead focus the analysis on the fine structure of the peak.
We do this because the light curve in the grey regions have features which may impact our measurements and are not likely to represent the ISCO crossing feature.

Our CNN does not have the ability to predict uncertainty from one light curve.
To circumvent this issue, we derive 10,000 light curves from our representative QSO 2237+0305 light curve by drawing values of $v_{\rm{eff}}$.
The length of each light curve was adjusted to be similar to light curves in the training sets in Section~\ref{machinelearning}.
Some light curves were clipped while others were extended depending on $v_{\rm{eff}}$.
Adjustments were made evenly at both sides and extensions were done by repeating the endpoint values with scatter based on the photometric uncertainty.
A resampled light curve is shown in Fig.~\ref{Figure14}.

\begin{figure}
    \centering
    \includegraphics[scale=0.5]{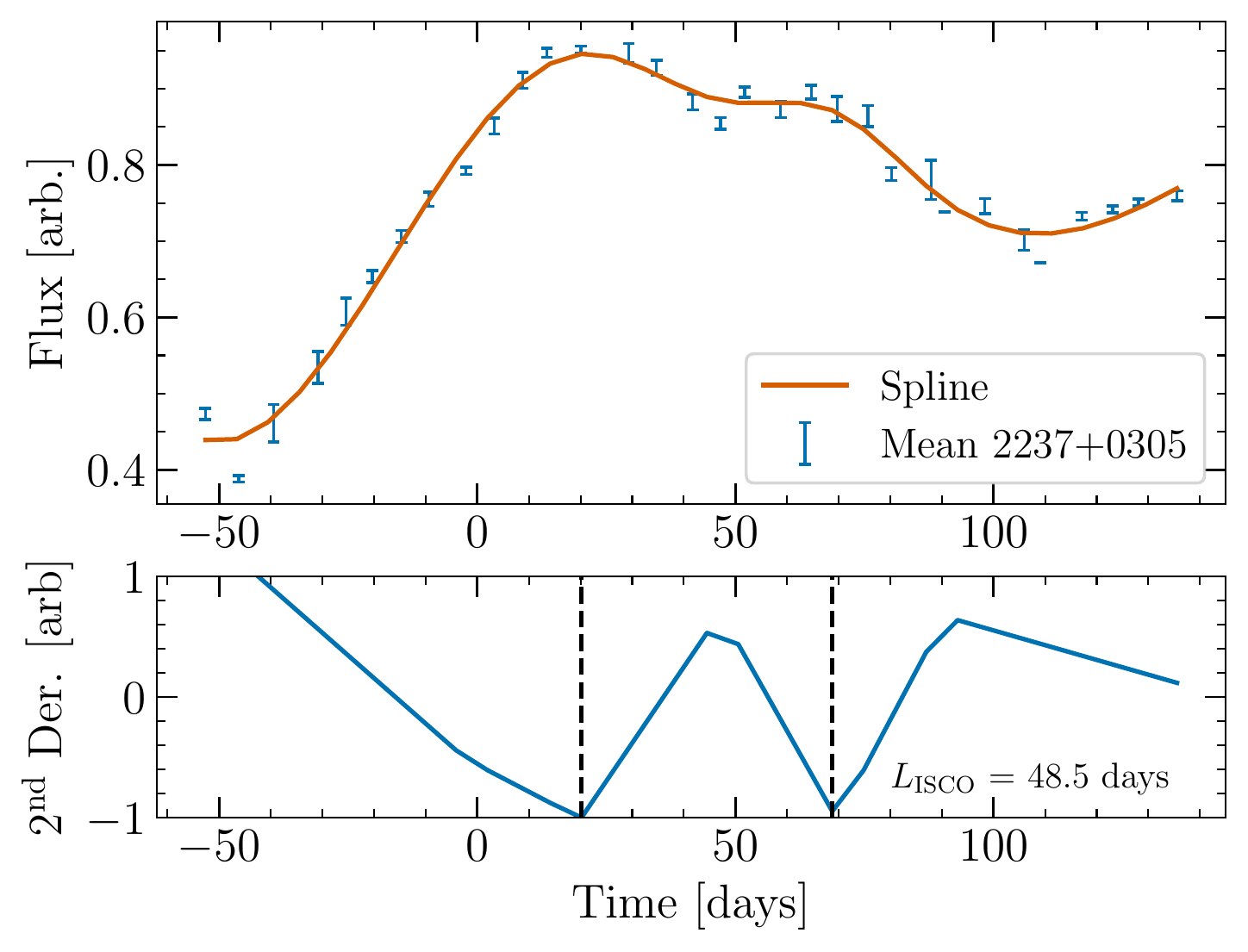}
    \caption{Top: Representative caustic-crossing event of QSO 2237+0305 plotted with a spline fit. The fitting algorithm converged with minimal noise tolerance. Bottom: The second derivative of a successful fit. $L_{\rm{ISCO}}$ is represented by the distance between dashed black bars.}
    \label{Figure15}
\end{figure}

\begin{figure}
    \centering
    \includegraphics[scale=0.5]{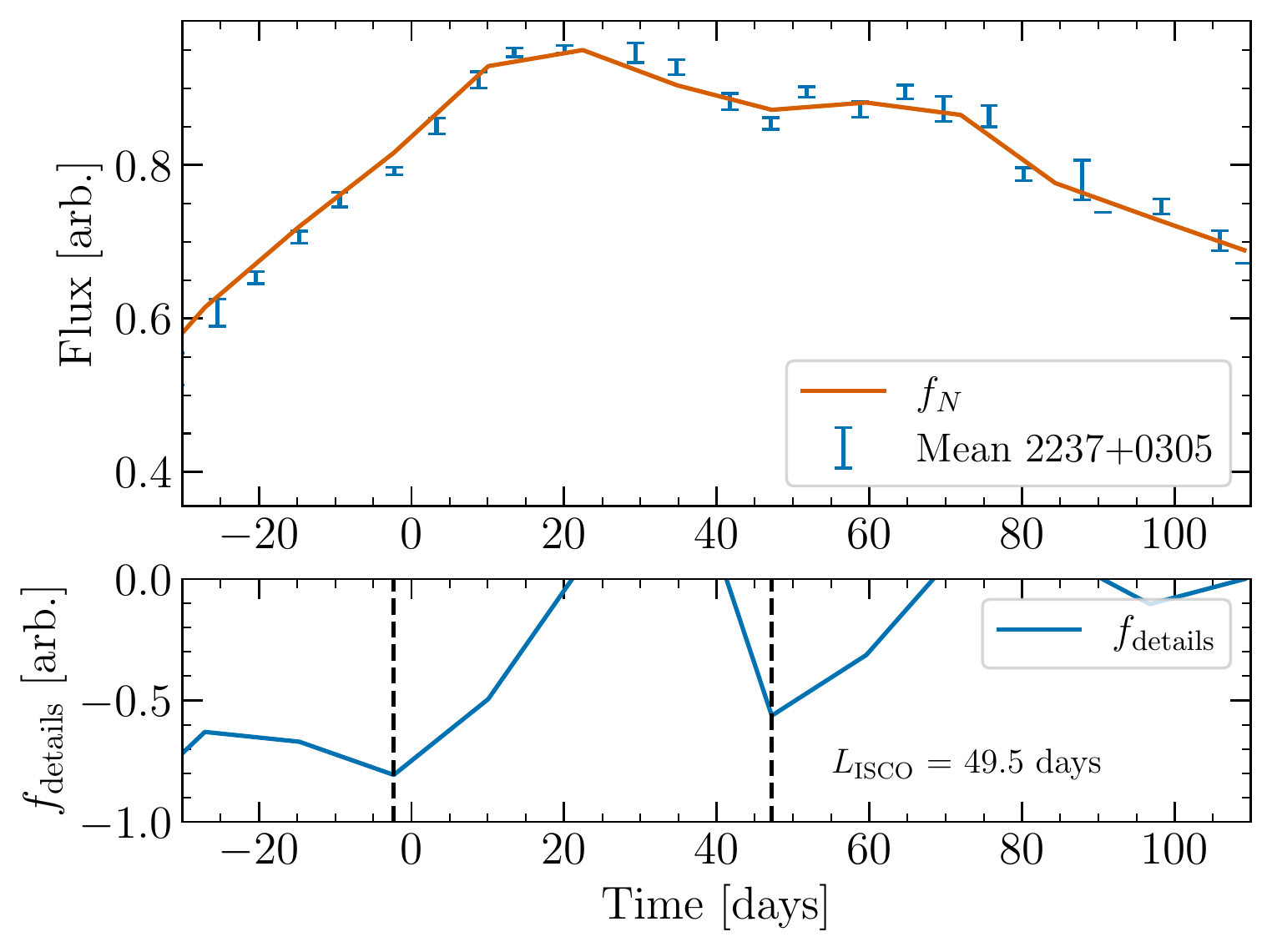}
    \caption{Wavelet analysis of representative caustic-crossing event of QSO 2237+0305. The measured ISCO crossing event is represented by the distance between the black bars.}
    \label{Figure16}
\end{figure}

Fig.~\ref{Figure15} presents our recovery of the ISCO crossing time as 48.5 days using the second derivative method.
Combined with our $R_{\rm{E}}$ and effective velocity model, we find $L_{\rm{ISCO}}$ = (1.4 $\pm$ 0.9) $\times 10^{13}$ m.
This is equivalent to the ISCO expected for a black hole with mass $(1.5 \pm 1.2) \times 10^{9} M_{\odot}$.

The wavelet method measures the ISCO crossing time as 49.5 days and is shown in Fig.~\ref{Figure16}.
We find the minima in $f_{\rm{details}}$ sooner than in the second derivative method but with similar separation as described in Section~\ref{sec:wavelet}.
This crossing time corresponds to $L_{\rm{ISCO}}$ = (1.4 $\pm$ 0.9) $\times 10^{13}$ m.
This is equivalent to the ISCO expected for a black hole with mass $(1.5 \pm 1.3) \times 10^{9} M_{\odot}$ based on our model assumptions.

We applied our trained CNN to 10,000 representations of the QSO 2237+0305 light curve like the one showed in Fig.~\ref{Figure14} to estimate the mass and inclination angle with their uncertainties.
Our CNN inferred log$_{10}(M_{\rm{BH}} / M_{\odot}) = (8.35 \pm 0.30)$ and the inclination angle was determined to be $45 \pm 23\degree$.
We did not attempt to estimate the impact angle as the data were derived from the combination of three separate caustic-crossing events.
We highlight that the network is able to place constraints on the mass and inclination even when trained on simulated light curves.

\section{Discussion and conclusions}
\label{Conclusionsection}

We have simulated a quasar accretion disc model which accounts for the strong lensing taken place by the central supermassive black hole.
With this model, we simulated caustic-crossing microlensing events using a high-resolution magnification map created through inverse ray-tracing.
We then applied realistic noise assuming high signal-to-noise optical follow-up from an 8-meter class telescope on quasars predicted to be discovered in LSST.
Our microlensing simulations assumed an isolated caustic structure.
While the isolated caustic case may be idealized, we note that some such events are expected to be observed in the hundreds to thousands of caustic-crossing events to be detected by LSST \citep{Neira20}.

For face-on accretion discs where relativistic effects are nearly insignificant, the ISCO imprints itself as a $\sim$1 to 10 per cent dip in brightness as it crosses the caustic. 
This dip occurred due to the dark ISCO crossing the caustic, where we lost a significant amount of flux from the disc.
For a moderately to highly inclined accretion disc, relativistic effects lead to strongly asymmetric peaks around the ISCO crossing and have the potential to distort the expected ISCO crossing signature. 
We have found that the distortion of the characteristic double-peak structure may lead to difficulty in measuring the length of the ISCO.

By creating an accretion disc image using ray-traced geodesics near the central SMBH, we find the projected ISCO length is a non-trivial function of several parameters.
The ISCO size scales with $M_{\rm{BH}}$ for any fixed inclination and impact angle.
The projected ISCO length depends on the inclination angle $i$, but the curving of geodesics reaching the far side of the accretion disc reduces the ISCO compression significantly.
We explored inclination angles up to $80\degree$ and found the compression ratio between the minor axis and major axis of the $L_{\rm{ISCO}}$ to be $\sim$ 70 per cent.
The compression of the minor axis due to inclination leads us to the range of projected ISCO crossing lengths, from which the impact angle $\phi$ then determines $L_{\rm{ISCO}}$.
We did not explore the degeneracy between mass and spin in simulations, and instead focused on a spinless black hole.

From these simulated light curves, we developed three methods in order to analyse these light curves.
Two methods focused on locating the beginning and end of the ISCO crossing event, while the third method used a CNN to infer simulation parameters from the light curve directly.
These methods were found to be successful for a wide range of the simulated parameter space.

The first method of measuring the projected ISCO we explored is the second derivative method, where a spline was fit to simulated light curves and the second derivative of this spline was analysed.
This method aimed to fit the original simulated light curve prior to the added noise in such a way the ISCO crossing signature was detectable.
This was done in order to calculate the distance between inflection points of the spline fit that captured the ISCO crossing event.
$L_{\rm{ISCO}}$ was found to be accurately recovered for many ISCO crossing events and simulation geometries.

The second method presented is the wavelet method, in which we decompose the simulated light curves into a set of discrete wavelet basis vectors.
We used the Daubechies basis to separate the long timescale features from the short timescale features such as the ISCO crossing.
The wavelet method was found to recover ISCO crossing features with similar ability to the second derivative method.
When we applied this to entire simulated populations, this method was less precise than the second derivative method.

We designed and trained a CNN to predict the ISCO length from the simulated light curves. 
The results of the validation set were found to show no substantial bias.
Unlike the other methods, the CNN had the ability to infer $L_{\rm{ISCO}}$ when one peak was suppressed or the noise was comparable to the ISCO signature.
The CNN also had much smaller uncertainties associated with predictions of $L_{\rm{ISCO}}$ compared to the other methods for simulated light curves.

We explored how the CNN was able to predict parameters of the system beyond $L_{\rm{ISCO}}$.
These parameters were the black hole mass $M_{\rm BH}$, the inclination angle of the accretion disc $i$, and the impact angle between the disc and the caustic $\phi$.
Furthermore, these light curves included a variety of signal-to-noise levels which were greater than our noise model as outlined in Appendix~\ref{IncreasedNoiseAppendix}.
We have found that black hole mass and inclination angle were predicted very well, with no significant biases and minor uncertainties even with these elevated levels of error.
The impact angle was predicted without significant bias, but had a larger uncertainty. 
We note that this impact angle only affects the orientation of the accretion disc with respect to the caustic feature and is not a physical parameter of the accretion disc.

In an attempt to understand what features were important to the CNN, two alternative simulations were performed. 
These sets represented deviations from an accretion disc with fixed $M_{\rm{BH}} = 10^{8} M_{\odot}$.
Our first alternative set had fixed accretion disc sizes and artificially adjusted ISCO sizes.
Our second alternative set had fixed ISCO sizes and artificially adjusted accretion disc sizes.
We found that predictions of the mass label on both of these alternative sets had negligible biases but increased uncertainty compared to the baseline.

We estimated $M_{\rm{BH}}$ for QSO 2237+0305 using archival data of caustic-crossing events with each method.
The second derivative and wavelet methods led us to an ISCO crossing time of 48.5 and 49.5 days, respectively.
These measurements are consistent with the time separation between peaks within the fine structure, measured by~\citet{Mediavilla15b} found to be $\sim$ 50 days.
Based on our models, these measured time separations correspond to $M_{\rm{BH}} = (1.5 \pm 1.2)$ and $(1.5 \pm 1.3) \times 10^{9} M_{\odot}$ for each method respectively.
The CNN estimated log$_{10}(M_{\rm{BH}} / M_{\odot}) = (8.35 \pm 0.30)$ after being trained on our simulated light curves.
Each of our methods are consistent with the range of estimates for QSO 2237+0305's SMBH mass of $1.2 \times 10^{9} M_{\odot}$~\citep{Assef11}, $2.0 \times 10^{8} M_{\odot}$~\citep{Sluse11}, and $9.0 \times 10^{7} M_{\odot}$~\citep{Hutsemekers21} based on BLR and microlensing measurements.

Beyond the mass, the CNN estimated the inclination of QSO 2237+0305 to be $45 \pm 23 \degree$.
It has been shown that high inclinations tend to be required ($i \geq 70 \degree$) for Kerr models of the accretion disc for QSO 2237+0305 by~\citet{Abolmasov12a}.
~\citet{Mediavilla15b} find that an inclination of up to $73 \degree$ is allowed.
On the other hand, ~\citet{Poindexter10a} find $i \leq 45 \degree$ at the 1 $\sigma$ level and~\citet{Hutsemekers21} find $i \sim 40 \degree$ which is consistent with our findings.
It is interesting that our CNN predicts a moderate inclination that is consistent with BLR studies while using a relativistic accretion disc model.
However, the wide range of uncertainty means the CNN was unable to place strong constraints on the inclination of QSO 2237+0305 from our simulated training set.
Further improvement to our estimates could come from exploring a wide range of microlens masses and training on a data set created with black hole spin information.

We have simulated high signal-to-noise light curves of an accretion disc caustic-crossing event and have managed to extract information regarding the ISCO size of the quasar.
We developed two direct methods which had the ability to extract the manifestation of the dark ISCO in caustic-crossing events for a wide range of parameters.
We demonstrated that a CNN could be used to predict $L_{\rm{ISCO}}$, as well as the mass, inclination, and orientation.
Each of these methods were applied to caustic-crossing events found in archival QSO 2237+0305 data and successfully estimated $M_{\rm{BH}}$.
This is an exciting era for microlensing studies, where wide-field surveys will regularly monitor and trigger alerts for high cadence follow-up of caustic-crossing events.

\section*{Data Availability}

Data of the QSO 2237+0305 light curves are publicly available by the OGLE\footnote{\url{https://ogle.astrouw.edu.pl}} and the GLITP\footnote{\url{https://vela.astro.uliege.be/themes/extragal/gravlens/bibdat/engl/lc_2237.html}} collaborations.
The data utilized in this paper are available upon reasonable request from the corresponding author. 
All simulated light curves within this study were created using the public codes \texttt{GPU-D}\footnote{\url{https://gerlumph.swin.edu.au/software/}} and \texttt{GYOTO}\footnote{\url{https://gyoto.obspm.fr}}.

\section*{Acknowledgements}
This project is supported by Schmidt Sciences, LLC.
We thank E. Mediavilla, J. Jim\'{e}nez Vincente, J. A. Mu\~{n}oz, and T. Mediavilla for sharing reduced data of QSO 2237+0305 caustic-crossing events. 
We are grateful to OGLE and GLITP collaborations for sharing data.
We thank Favio Neira for assistance with the velocity model used for the analysis of QSO 2237+0305.
We thank the scientific editor and anonymous referee who have provided comments which have greatly strengthened this work.
Simulated error is based on observations obtained at the international Gemini Observatory, a program of NSF’s NOIRLab, which is managed by the Association of Universities for Research in Astronomy (AURA) under a cooperative agreement with the National Science Foundation on behalf of the Gemini Observatory partnership: the National Science Foundation (United States), National Research Council (Canada), Agencia Nacional de Investigaci\'{o}n y Desarrollo (Chile), Ministerio de Ciencia, Tecnolog\'{i}a e Innovaci\'{o}n (Argentina), Minist\'{e}rio da Ci\^{e}ncia, Tecnologia, Inova\c{c}\~{o}es e Comunica\c{c}\~{o}es (Brazil), and Korea Astronomy and Space Science Institute (Republic of Korea).

In addition to user created code, the following Python modules were also used: \texttt{Numpy}\footnote{\url{https://numpy.org}}, \texttt{Scipy}\footnote{\url{https://scipy.org}}, \texttt{Astropy}\footnote{\url{https://www.astropy.org}}, \texttt{Matplotlib}\footnote{\url{https://matplotlib.org}}, \texttt{Tensorflow/Keras}\footnote{\url{https://www.tensorflow.org}}, and \texttt{PyWavelets}\footnote{\url{https://github.com/PyWavelets/pywt}}.
These resources have been invaluable to both creating and exploring these simulations.

\bibliographystyle{mnras}
\bibliography{bib.bib}

\appendix

\section{Increased photometric uncertainty for CNN generalization}
\renewcommand{\thefigure}{A.\arabic{figure}}
\renewcommand{\thetable}{A.\arabic{table}}
\setcounter{figure}{0}  
\setcounter{table}{0} 

\begin{figure*}
    \centering
    \includegraphics[width=0.97\textwidth]{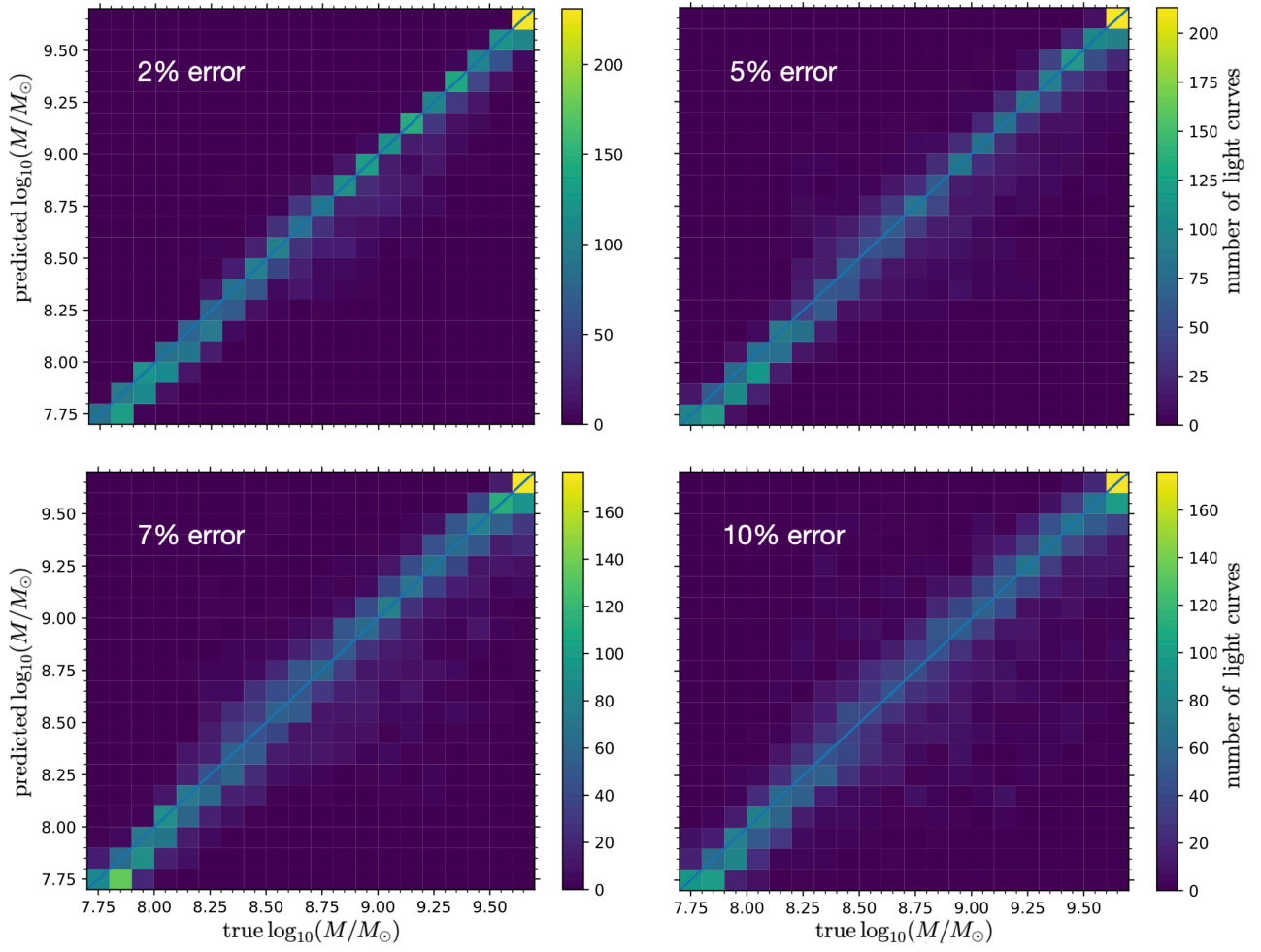}
    \caption{Predicted values vs true values for the black hole mass in data sets with increased photometric error levels ranging from 2 to 10 per cent. The RMSE for the mass predictions are given in table~\ref{TableA1}.}
    \label{figureA1}
\end{figure*}

\label{IncreasedNoiseAppendix}
We tested the generalization of the CNN to make predictions at more realistic noise levels. 
In order to do so, we apply additional photometric noise at 2, 5, 7, and 10 per cent levels and attempt to infer the black hole mass used in the simulation.
Results of this test are shown in Fig.~\ref{figureA1} where the root mean squared error (RMSE) for mass predictions are given in Table~\ref{TableA1}.
This metric is the square root of the loss function used to train the network, and combines the bias and uncertainty into one value.
These inferences continued to be more accurate and precise than the second derivative method and wavelet methods.
Due to this result, our targeted sets included greater photometric noise to generalize the CNN to more realistic observations.
Having a more generalized network was also important when applied to the observational data of QSO 2237+0305.

\begin{table}
    \centering
    \caption{Root mean squared error for mass prediction on the training set with increased photometric noise levels.}    
    \begin{tabular}{lc}
    Increased error level & RMSE for mass [dex] \\
    \hline
    $2\%$ & 0.10 \\
    $5\%$ & 0.16 \\
    $7\%$ & 0.20 \\
    $10\%$ & 0.24 \\
    \end{tabular}
    \label{TableA1}
\end{table}

\bsp	
\label{lastpage}

\end{document}